\documentclass[preprint]{JHEP3} % 10pt is ignored!

\JHEPspecialurl{http://jhep.sissa.it/JOURNAL/JHEP3.tar.gz}
\usepackage{graphicx}
\usepackage{amsmath,amssymb}
\usepackage{slashed}

\newcommand\fverb{\setbox\fverbbox=\hbox\bgroup\verb}
\newcommand\fverbdo{\egroup\medskip\noindent%
			\fbox{\unhbox\fverbbox}\ }
\newcommand\fverbit{\egroup\item[\fbox{\unhbox\fverbbox}]}
\newbox\fverbbox

\newcommand{\e}{\epsilon}

\newcommand{\ba}{\begin{equation}}
\newcommand{\ea}{\end{equation}}
\newcommand{\be}{\begin{eqnarray}}
\newcommand{\ee}{\end{eqnarray}}

\def\e{\epsilon}

\def\e{\epsilon}

\title{\boldmath NLO QCD corrections for $W^+W^-$ pair production in
  association with two jets at hadron colliders \unboldmath }

\author{Tom Melia \\Rudolf Peierls Centre for Theoretical Physics, 1
  Keble Road, University of
  Oxford, UK\\
  Email: \email{t.melia1@physics.ox.ac.uk} }

\author{Kirill~Melnikov \\Department of Physics and Astronomy, Johns
  Hopkins
  University, Baltimore, MD 21218, USA\\
  Email: \email{melnikov@pha.jhu.edu} }

\author{Raoul R\"ontsch \\Rudolf Peierls Centre for Theoretical
  Physics, 1 Keble Road, University of
  Oxford, UK\\
  Email: \email{r.rontsch1@physics.ox.ac.uk} } 

\author{Giulia Zanderighi \\Rudolf Peierls Centre for Theoretical
  Physics, 1 Keble Road, University of
  Oxford, UK\\
  Email: \email{g.zanderighi1@physics.ox.ac.uk} }

\received{\today} 		%%
\accepted{\today}		%% These are for published papers.

\abstract{We compute the NLO QCD corrections to the pair production of
  $W$-bosons in association with two jets at the Tevatron and the LHC.
  This process is an important background to heavy Higgs-boson
  production in association with two jets, either in gluon or weak
  boson fusion.  We consider leptonic decays of $W$-bosons and include
  all the spin correlations exactly. For natural choices of the
  renormalization scale, the NLO QCD corrections to $pp(\bar p) \to
  W^+W^-jj$ are moderate but different for various values of the
  center-of-mass collision energy at the LHC and the Tevatron,
  emphasizing the need to compute them explicitly.  }


\begin{document} 

\section{Introduction}

The Large Hadron Collider (LHC) has begun to explore the Standard
Model (SM) of particle physics in a new energy regime and will in time
gather more data than any previous hadron collider experiment. To
further our understanding of the SM and as to what may lie
beyond it, we attempt to describe outcomes of proton collisions in
sufficient detail, for comparison with the observed data.  An accurate
knowledge of SM processes is particularly important as
their cross-sections are often much larger than those for
many interesting New Physics processes.  Unless physics beyond the
SM presents itself in a stark way, disentangling it from
SM backgrounds will require an accurate description of the
latter.  Parton level calculations at leading order (LO) in the strong
coupling constant are often insufficient for this purpose.  They
exhibit a strong unphysical dependence on factorization and
renormalization scales, leading to large uncertainties in the
predictions.
Data-driven estimates of the backgrounds are also subject to large
uncertainties if they rely on LO theoretical predictions: here the
idea is to determine the normalization of the LO cross-section for a
given background process in a region essentially free from any New
Physics signal. Once the LO is ``validated'' using data, one
extrapolates it to the region of interest. It is clear that such a
procedure can only work if higher order QCD corrections are uniform
over phase space, which is not guaranteed in general.
As follows from many successful analyses at the Tevatron, a good way
to reduce the uncertainty is to extend the theoretical description of
a given process to next-to-leading order (NLO) in perturbative QCD.

The past five years have seen an extraordinary progress in the
development of methods that are suitable to deal with NLO QCD
computations for high-multiplicity processes. Refinements of
traditional computational techniques based on the Passarino-Veltman
reduction of tensor integrals led to the development of
highly-efficient, Feynman-diagram-based technology for NLO QCD
computations \cite{Denner:2002ii,Binoth:2005ff,Denner:2005nn}.  At the
same time, new techniques based on unitarity and on-shell methods
\cite{Britto:2004nc,Britto:2004tx,Ossola:2006us,Forde:2007mi,Ellis:2007br,Giele:2008ve,
  Ossola:2008xq} sufficiently matured to become relevant for practical
applications.  As a result, a large number of $2 \to 4$ processes were
studied at NLO in QCD in the past two years.  The list includes $pp
\rightarrow W (Z,\gamma) + 3$ jets \cite{Berger:2009ep,
  Berger:2009zg,KeithEllis:2009bu,Ellis:2009zw,Melnikov:2009wh,Berger:2010vm},
$pp \rightarrow t\bar{t} b \bar{b}$ \cite{Bredenstein:2009aj,
  Bredenstein:2010rs,Bevilacqua:2009zn}, $pp \rightarrow t \bar{t} +
2$ jets \cite{Bevilacqua:2010ve}, $pp \rightarrow b \bar{b} b \bar{b}$
\cite{Binoth:2009rv}, $p p \rightarrow t\bar{t} \rightarrow W^+W^-b
\bar{b}$ \cite{Bevilacqua:2010qb,Denner:2010jp}, and $pp \rightarrow
W^+W^+ + 2$ jets \cite{Melia:2010bm}.  This last process has been
implemented recently in POWHEG-BOX~\cite{Melia:2011gk}. This combines
NLO accuracy with a parton shower detailed description of the final
state.

The first $
2\rightarrow 5$ process, $pp \rightarrow W + 4$ jets, has also been
computed recently through NLO QCD using on-shell methods
\cite{Berger:2010zx}. Some groups also started employing those
advances with the intention of developing a platform for fully
automated NLO QCD calculation \cite{Mastrolia:2010nb,Hirschi:2011pa}.

The goal of this paper is to study the production of a pair of
$W$-bosons in association with two jets in hadron collisions,
including the NLO QCD corrections. The production of a $W$-boson pair
in association with zero, one or two jets is an important background
to searches for intermediate and heavy Higgs boson, where the decay $H
\to W^+W^-$ opens up.  At the Tevatron, searches for 
intermediate-mass Higgs bosons treat processes $p \bar p \to H+n~{\rm
  jets}$, $n=0,1, \ge 2$ separately, because dominant backgrounds
depend on the number of identified jets in the final state (see
e.g. \cite{higgscdf}).  While most of the sensitivity in Higgs-boson
searches comes from the process with the largest cross-section, $p
\bar p \to H + 0~{\rm jets}$, the production of the Higgs boson in
association with two jets is also relevant
\cite{Campbell:2006xx,Anastasiou:2009bt}.  Because $p \bar p \to
W^+W^- jj$ is an irreducible background to the Higgs-boson production
in association with two jets, it is important to have NLO QCD
predictions for this process.

There is yet another reason to want an improved description of
$W^+W^-jj$ production in hadron collisions. At the LHC the Higgs boson
can be produced with a sizable cross-section in weak boson fusion
(WBF) \cite{Campbell:2006xx,Berger:2004pca}.  In addition to the
Higgs-boson decay products which, as we assume, are pairs of
$W$-bosons, the signature of the process involves two forward tagging
jets.  In this case, $p p\rightarrow W^+W^-jj$ is the irreducible
background.  The Higgs-production cross-section in WBF is known
through NLO QCD \cite{Berger:2004pca,Han:1992hr,Figy:2003nv}, and it
is desirable for the dominant background process to be known to the
same order in perturbative QCD as well.

Finally, we note that jets, charged leptons and missing energy is one
of the classic signatures of dark-matter-type processes at
colliders. In such scenarios the missing energy appears due to the
dark-matter candidate escaping the detector. The process
$pp\rightarrow W^+W^-jj$ is a SM background with a similar
signature, where leptonic decays of $W$-bosons lead to invisible
neutrinos.

Several studies in the past decade addressed the production of
$W$-boson pairs in hadron collisions, including NLO QCD corrections.
In particular, $pp\rightarrow W^+W^-$ with no jets was studied in
Refs.~\cite{Dixon:1999di,Campbell:1999ah,Ohnemus:1991gb}. The
production of a pair of $W$-bosons in association with one jet
including decays to leptons was studied through NLO QCD in
Refs.~\cite{Campbell:2007ev,Dittmaier:2009un}. In both cases, for the
choice of the renormalization and factorization scales $\mu = M_W$,
QCD corrections were found to be significant, of the order of
$(25-50)\%$.  These results further motivate the need to understand the
production of $W^+W^-$ in association with two or even more jets at
NLO in QCD.

In this paper we allow for leptonic decays of the $W$-pair including
all spin correlations.  Dilepton final states are the ones that are
relevant for ongoing Higgs searches at the Tevatron and, in general,
these final states provide the cleanest signature to identify the
production of $W$-bosons at a hadron collider. For this reason, we
find it reasonable to focus on these states only.

We remind the reader that the branching fraction for the $W$-boson to
decay to a definite-flavor lepton final state is about ten
percent. Since we have two $W$-bosons decaying leptonically, we get a
hit by a factor ${\cal O}(10^{-2})$ when the cross-section for
dilepton final state is compared with the cross-section for stable
$W$-bosons.  It is therefore amazing that the cross-section for the
process $pp \to (W^+ \to \nu_{\mu} \mu^+) +(W^- \to e^- \bar{\nu}_e)
+jj$ is still reasonably large.  In particular, we find that the cross
section for the LHC running at an energy of 7 TeV is around $40~{\rm
  fb}$, which means that a few of these events should have already
been seen at this collider at the time of publication, and quite a
significant number of such events should be produced at the LHC by the
end of the next year.  The cross section further increases to about
0.14 pb at 14 TeV, so there is no doubt that the experimental study of
this process is feasible. Even at the Tevatron, where the
cross-section with the ``Higgs-like'' cuts for one flavor assignment
is just $2.0~{\rm fb}$, assuming fifty percent efficiency, about $40$
$e^+e^- \bar \nu \nu jj$, $\mu^+\mu^- \bar \nu \nu jj$, $\mu^+e^- \bar
\nu \nu jj$, $e^+\mu^- \bar \nu \nu jj$ events should have been
recorded already.

The computation of NLO QCD corrections to hadro-production of $W^+W^-$,
in association with two jets, is also interesting from the point of
view of further developing on-shell methods for one-loop computations.
Recall that, as currently formulated, on-shell methods require
ordering of external lines which is achieved by working with
color-ordered \cite{Mangano:1990by} or primitive amplitudes
\cite{Bern:1994fz,Bern:1996je,Bern:1997sc}.  These techniques work
best if all external particles carry color charges, while their
implementation becomes more involved as the number of colorless
particles in the process increases.
The only process which involves {\em two} colorless particles that has
been computed with unitarity methods before, $pp\to W^+W^+ jj$
\cite{Melia:2010bm}, is simpler than the calculation presented here
since, among other things, only a smaller number of sub-processes
contribute.  As explained in the next Section, the presence of two
colorless particles in the final state whose total electric charge is
zero provides some additional difficulty. Nevertheless, it is possible
to handle these complications with on-shell methods.

The remainder of the paper is organized as follows. In
Section~\ref{tech}, we provide technical details of the
calculation. In Section~\ref{pheno} we discuss phenomenological
results for the QCD production of $W^+W^-jj$, with leptonic decays of
the $W$-bosons, at the Tevatron and the LHC. We conclude in Section 4.
We provide numerical results for various one-loop primitive
amplitudes, as well as squared amplitudes summed over helicity and
color for $(W^+ \to  \nu_{\mu} \mu^+) +(W^- \to e^- \bar{\nu}_e) jj$
hadronic production in the Appendix.

\section{Technical Details}
\label{tech}
In this Section we present technical details specific to this
calculation.
Within a subtraction formalism, a NLO calculation involves three
components: virtual corrections, real emission corrections, and
subtraction terms for soft and collinear divergences. The virtual
corrections are computed using $D$-dimensional generalized unitarity
\cite{Ellis:2007br,Giele:2008ve}.
A detailed description of the implementation of this method can be
found in \cite{Ellis:2008qc}; we have followed this implementation,
modifying and extending it to deal with the presence of an additional
$W$-boson in the final state.

The full one-loop amplitude can be built by summing products of
color-ordered partial amplitudes
\cite{Bern:1994fz,Bern:1996je,Bern:1997sc} over all permutations of
the colored particles, with appropriate color factors. The partial
amplitudes are further decomposed into primitive amplitudes. The
ordering of all particles with color charges is fixed in primitive
amplitudes.
$D$-dimensional unitarity cuts reduce one-loop primitive amplitudes to
linear combinations of products of tree-level helicity amplitudes,
which are computed using Berends-Giele recursion relations
\cite{Berends:1987me}. These relations are also used to compute
tree-level amplitudes which are required for calculations of LO
cross-sections, real emission corrections and subtraction terms for
soft and collinear emissions.  We implement subtraction terms
following the Catani-Seymour procedure \cite{Catani:1996vz}, with the
$\alpha$-parameter optimization as described in
Ref.~\cite{Nagy:1998bb,Nagy:2003tz}.  We embed our calculations within
the framework of the {\sf MCFM} program \cite{Campbell:2000bg} and use
the {\sf QCDloop} program to calculate the scalar one-loop integrals
\cite{Ellis:2007qk}.

Since only color-charged particles are ordered in primitive
amplitudes, all possible insertions of the $W$-bosons must be
considered when tree-level or one-loop primitive amplitudes are
computed. While this implies a certain amount of non-trivial
book-keeping in the construction of a numerical program, this can be
done without much trouble.  The real problem, however, is that cuts of
different parent diagrams must be combined in certain cases to produce
gauge-invariant tree-level amplitudes in the context of unitarity
cuts. This implies that different parent diagrams can not be treated
independently and this creates considerable overhead.  Furthermore, we
must include the possibility of the $W^+W^-$ pair being produced via
an intermediate neutral vector boson, such as an off-shell $\gamma$ or
$Z$.

In this calculation, we do not consider the production of top quarks
in the final state, as these are processes with a distinct
experimental signature. Furthermore, we neglect top-quark
contributions in virtual diagrams and treat all other quarks as
massless.  Since top quarks in virtual diagrams originate from $b \to
W t^* $ transitions, we decided to completely exclude bottom quarks in
our calculation as well. This is a reasonable approximation since the
$b$-content of the proton is subdominant both at the Tevatron and the
LHC.  We also neglect $g^* \to b \bar b$ splitting, both real and
virtual.  We believe that this effect is also quite small as can be
seen from the $b$-quark contribution to the QCD $\beta$-function,
relative to contributions of gluons and four other quarks.  Although
we do not expect that the complete omission of quarks in the third
generation impacts our results in any significant way, we hope to
include them in the calculation in the future. The framework to do so,
within the generalized D-dimensional unitarity approach, has already
been fully elaborated in Refs.~\cite{Ellis:2008ir,Melnikov:2010iu}.
Before continuing, we point out that in this paper we do not include
contributions from one-loop diagrams where the $\gamma/Z$ or the
$W^+W^-$ pair couple directly to a loop of virtual quarks, creating a
diagram of the ``light-by-light scattering'' type. These diagrams form
a finite, gauge-invariant class of amplitudes that can be dealt with
separately. In particular, amplitudes for the partonic process $gg \to
W^+W^- gg$ which does not appear at tree-level also belong to that
class of amplitudes. As pointed out in Ref.~\cite{Binoth:2006mf},
processes of that type may be quite important because of the large
gluon flux at the LHC. We plan to return to the discussion of the
amplitudes where $W^+W^-$ pair couples directly to a closed quark loop in a
separate publication.
Finally, in this calculation we neglect mixing between up and down
quarks of different generations and set the CKM matrix to the identity
matrix.

The production of $W^+W^- jj$ can occur through both electroweak and
QCD mechanisms. The NLO QCD corrections for the electroweak production
have already been calculated in Ref. \cite{Jager:2006zc}. While these
mechanisms can interfere even at leading order, these interference
terms are strongly suppressed. First, at partonic level, the
electroweak production of $W^+W^- jj$ involves four quarks. However,
given the large gluon luminosity at the LHC, four-quark contributions
to $W^+W^-jj$ production cross-section amount to only about
15\%. Moreover, the interference that occurs in a four-quark process
can only happen for certain combinations of quark flavors, and it is
color-suppressed. We therefore neglect this interference, and present
results for the QCD production alone.

In order to describe tree-level and one-loop virtual corrections to $pp
(p\bar p) \to W^+W^-jj$ we require partonic processes with either two
quarks and two gluons $0 \to \bar q_1 q_2 \;gg \; W^+W^-$ or with four
quarks $0 \to \bar q_1 q_2 \; \bar q_3 q_4 W^+W^-$.  Given the
difference in color- and flavor structures, we discuss these two
partonic processes separately in the next two Subsections.

\subsection{Processes with $W^+W^-$ pair, a quark pair and two gluons}
In this Section, we consider the partonic process $0 \to \bar q_1 q_2
gg W^+W^-$.  Since we neglect mixing between up and down quarks of
different generations, in two-quark amplitudes both quarks $q_{1,2}$
have the same flavor.

To obtain the full NLO cross-section, we need to consider all possible
crossings between the partons; the initial state partons -- as well as
the jets -- may be either gluons or quarks.  The tree level amplitude
for the process $0 \rightarrow (\bar{q} q) + (W^{+} \rightarrow
\nu_{\mu} + \mu^+) + (W^- \rightarrow e^- + \bar{\nu}_{e}) + g +g$ can
be written as
\begin{equation}
\begin{split}
 & \mathcal{A}^{\mathrm{tree}}(\bar{q}_1,q_2;,\nu_{\mu},\mu^+,e^-,\bar{\nu}_{e};
g_3,g_4) = g_s^2 \biggl( \frac{g_W}{\sqrt{2}} \biggr)^4 
P_{W}(s_{\nu_{\mu} \mu^+}) P_{W}(s_{e^{-} \bar{\nu}_e}) 
 \\
&\;\;\;\;\;\;\;\;\;\;\;\;
 \times \biggl((T^{a_3} T^{a_4})_{\bar{i}_1i_2} A_0(\bar{q}_1,q_2;g_3,g_4) + 
  (T^{a_4} T^{a_3})_{\bar{i}_1 i_2} A_0(\bar{q}_1,q_2;g_4,g_3) \biggr).
\end{split} \label{ggtree}
\end{equation}
In Eq.(\ref{ggtree}), $g_s$ and $g_W$ are the strong and weak coupling
constants, respectively, leptonic labels have been suppressed on the
right hand side, and $P_{W}$ are Breit-Wigner propagators
\begin{equation}
 P_W(s) = \frac{s}{s - M_W^2 + i\Gamma_W M_W}.
\end{equation}
with $s_{l \nu} = (p_l + p_{\nu})^2$. In addition, $M_W$ and
$\Gamma_W$ are the $W$-boson mass and width, and the generators of the
$SU(3)$ color group are normalised to $\mathrm{Tr}(T^a T^b) =
\delta^{ab}$.  In Eq.(\ref{ggtree}), $A_0$ denote the color-ordered
amplitudes.  The flavor of the quark line fixes the electric charges
of $q_1$ and $q_2$ and, simultaneously, the ordering of $W^+$ and
$W^-$ along the quark line.  However, as we pointed out before, the
relative ordering of $W^\pm$ bosons and gluons is not fixed.
Additionally, we need to consider the possibility that $W$-bosons are
produced through an intermediate (off-shell) $Z$-boson or photon. Thus
we write
\begin{equation}
 A_0(\bar{q}_1,q_2;g_3,g_4)  = A_0^{(WW)}([\bar{q}_1,W,W,q_2];g_3,g_4) 
+ 
C^{(q_2,h_2)} A_0^{(\gamma/Z)}([\bar{q}_1,\gamma/Z,q_2];g_3,g_4),    \label{WWZgam}
\end{equation}
where the first term describes an amplitude where $W$-bosons couple
directly to the quark line and the second term describes an amplitude
where such coupling occurs through a $\gamma$ or $Z$. The factor
$C^{(q_2,h_2)}$ is given by
\begin{equation} 
C^{(q,h)} =  2 Q^{(q)}\sin^2 \theta_W + P_Z(s_Z) (T_3^{(h)} - 2Q^{(q)}\sin^2 \theta_W)\,,    \label{gamZ}
\end{equation}
where $Q^{(q)}$ and $h$ are the electromagnetic charge and helicity of
the quark $q$, $T_3^{(-)} = 1$ and $T_3^{(+)} = 0$, $\theta_W$ is the
weak mixing angle, and $s_Z = (p_{W^+} + p_{W^-})^2 = (p_{\nu_{\mu}} +
p_{\mu^+} + p_{e^-} + p_{\bar{\nu}_e} )^2 $.  Note that, 
because $W$-bosons only couple to left-handed quarks, 
 the first term
in Eq.~\eqref{WWZgam} is zero if the quark is right-handed. 
We account
for the decay $W^{\pm} \rightarrow l^{\pm}(p_l) + \nu_l(p_{\nu})$, by
using the $W^\pm$ polarization vectors constructed from lepton
spinors. For example, in case of the $W^+$ boson, the polarization
vector reads
\begin{equation}
 \epsilon^{\mu}_{-}(p_{\nu},p_{l^+}) = \frac{\bar{u}(p_{\nu}) \gamma^{\mu} \gamma_{-} v(p_{l^+})}{(p_{l^+}+p_{\nu})^2} ;  \hspace{5 mm} \gamma_{-} = \frac{1 - \gamma_5}{2}.
\end{equation}

\begin{figure}[t]
\begin{center}
\includegraphics[width=3.2cm]{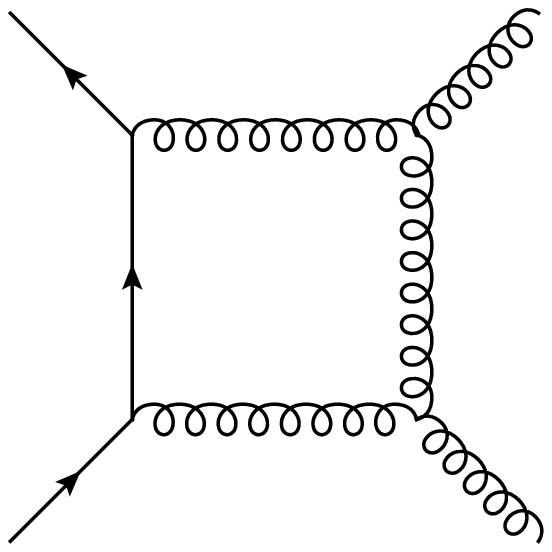} \hspace*{0.5cm}
\includegraphics[width=3.2cm]{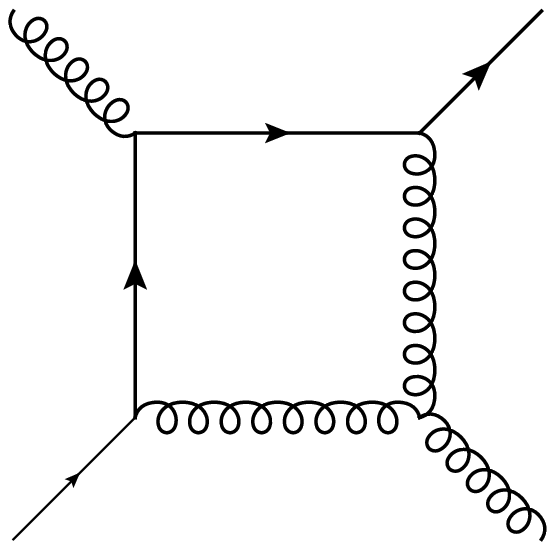} \hspace*{0.5cm}
%\vspace*{0.3cm}
\includegraphics[width=3.2cm]{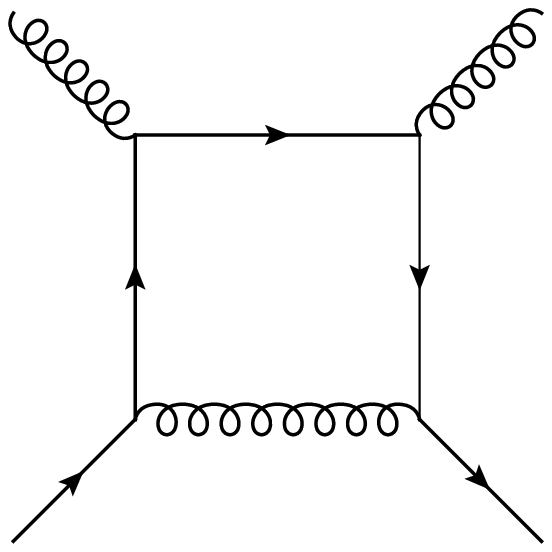} \hspace*{0.5cm}
\includegraphics[width=3.2cm]{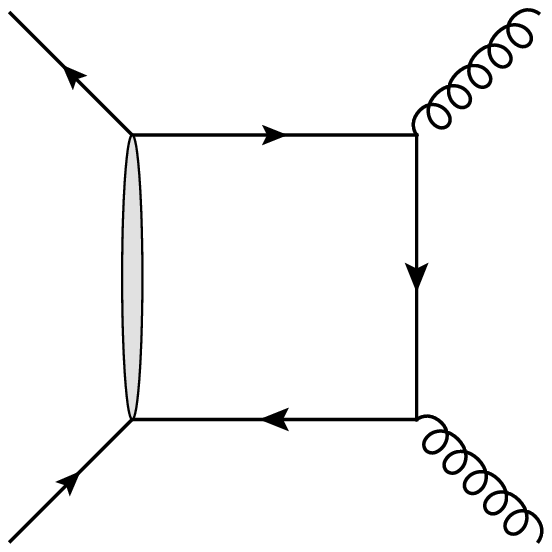}

\begin{picture}(0,0)(0,0)
\put(-180,0){(a)}
\put(-68,0){(b)}
\put(44,0){(c)}
\put(158,0){(d)}

\put(-220,8){$\bar q_1$}
\put(-130,8){$g_4$}
\put(-220,110){$q_2$}
\put(-130,110){$g_3$}

\put(-108,8){$\bar q_1$}
\put(-18,8){$g_4$}
\put(-108,110){$g_3$}
\put(-18,110){$q_2$}

\put(4,8){$\bar q_1$}
\put(94,8){$q_2$}
\put(4,110){$g_3$}
\put(94,110){$g_4$}

\put(116,8){$\bar q_1$}
\put(206,8){$g_4$}
\put(116,110){$q_2$}
\put(206,110){$g_3$}

\end{picture}

\caption{Primitive amplitudes a) $A_1(\bar{q}_1,q_2,g_3,g_4)$, 
b) $A_{1}(\bar{q}_1,g_3,q_2,g_4)$, c) $A_{1}(\bar{q}_1,g_3,g_4,q_2)$, and
d) $A_{1}^{[1/2]}(\bar{q}_1,q_2,g_3,g_4)$. $W$
bosons are not shown.} \label{fig:ggvirtampl}
\end{center}
\end{figure}

The computation of real emission contributions requires tree
amplitudes with an additional gluon in the final state. The color
decomposition reads
 \begin{equation}
 \begin{split}
&   \mathcal{A}^{\mathrm{tree}}(\bar{q}_1,q_2;\nu_{\mu},\mu^+,e^-,\bar{\nu}_{e};g_3,g_4,g_5) = g_s^3 \biggl( \frac{g_W}{\sqrt{2}} \biggr)^4 
P_{W}(s_{\nu_{\mu} \mu^+}) P_{W}(s_{e^{-} \bar{\nu}_e}) 
  \\
&\;\;\;\;\;\;\;\; \times 
\sum_{\sigma \in S_3} (T^{a_{\sigma_3}} T^{a_{\sigma_4}}
T^{a_{\sigma_5}})_{\bar{i}_1 i_2}
A_0(\bar{q}_1,q_2;g_{\sigma_3},g_{\sigma_4},g_{\sigma_5}), 
 \end{split}
 \end{equation}
where $S_i$ denotes the permutation of $i$ indices. 
The flavor/helicity properties of the amplitudes with three and two
gluons are identical and have been discussed for the two-gluon case.

The decomposition of the one-loop amplitudes in terms of left-handed
primitive amplitudes \cite{Bern:1994fz,Ellis:2008qc} reads
\begin{equation}
 \begin{split}
  & \mathcal{A}^{\mathrm{1L}}(\bar{q}_1,q_2;\nu_{\mu},\mu^+,e^-,\bar{\nu}_{e};g_3,g_4) = g_s^4 \biggl( \frac{g_W}{\sqrt{2}} \biggr)^4 
P_{W}(s_{\nu_{\mu} \mu^+}) P_{W}(s_{e^{-} \bar{\nu}_{e}}) 
\\
 &\;\;\;\;\;\; \times \sum_{\sigma \in S_2} \biggl[ \bigl(T^{x_2} 
T^{a_{\sigma_3}} T^{a_{\sigma_4}} 
T^{x_2} \bigr)_{\bar{i}_1 i_2} A_1(\bar{q}_1,g_{\sigma_4},g_{\sigma_3},q_2) 
\\
&\;\;\;\;\;\;\;\;\;\;\;
+ 
\bigl(T^{x_2} T^{a_{\sigma_3}} T^{x_1} \bigr)_{\bar{i}_1 i_2} \bigl(f^{a_{\sigma_4}}\bigr)_{x_1 x_2} A_1(\bar{q}_1,g_{\sigma_3},q_2,g_{\sigma_4})  \\ 
&\;\;\;\;\;\;\;\;\;\;\;
+
\bigl( T^{x_2} T^{x_1}\bigr)_{\bar{i}_1 i_2} \bigl(f^{a_{\sigma_3}} f^{a_{\sigma_4}} \bigr)_{x_1 x_2} A_1(\bar{q}_1,q_2,g_{\sigma_4},g_{\sigma_3})  \\
& \;\;\;\;\;\;\;\;\;\;\; + 
\frac{n_f}{N_c}  \mathrm{Gr}_4 A_1^{[1/2]}(\bar{q}_1,q_2,g_{\sigma_3},g_{\sigma_4})
\biggr].
 \end{split} \label{ggvirt}
\end{equation}
In Eq.(\ref{ggvirt}) we introduced the color factor 
\begin{equation}
\mathrm{Gr}_4 = N_c \bigl(T^{a_{\sigma_3}} T^{a_{\sigma_4}} \bigr)_{\bar{i}_1 i_2} - 
\mathrm{Tr}\bigl(T^{a_{\sigma_3}} T^{a_{\sigma_4}} \bigr)
\delta_{\bar{i}_1 i_2}\,. 
\end{equation}
We build up the virtual amplitude from eight primitive amplitudes:
$A_1(\bar{q}_1,g_4,g_3,q_2)$, $A_{1}(\bar{q}_1,g_3,q_2,g_4)$,
$A_{1}(\bar{q}_1,q_2,g_4,g_3)$ and
$A_{1}^{[1/2]}(\bar{q}_1,q_2,g_3,g_4)$, shown in
Fig.~\ref{fig:ggvirtampl} and another four amplitudes, obtained by
swapping the gluons $g_3 \leftrightarrow g_4$. In
Fig.~\ref{fig:ggvirtampl}, we introduce a `dummy line' for the
primitive amplitude $A_{1}^{[1/2]}(\bar{q}_1,q_2,g_3,g_4)$. This
allows us to draw this primitive amplitude -- which has the external
gluons attached to a fermion loop -- as formally having six
loop-momentum-dependent propagators\footnote{We count external
  $W$-bosons, which are not shown in Fig.~\ref{fig:ggvirtampl}.}.  The
$W$-bosons couple to the dummy lines, but dummy lines cannot be cut.

\subsection{Processes with $W^+W^-$ and two quark pairs}

\begin{figure}[t]
\begin{center}
\includegraphics[width=8cm]{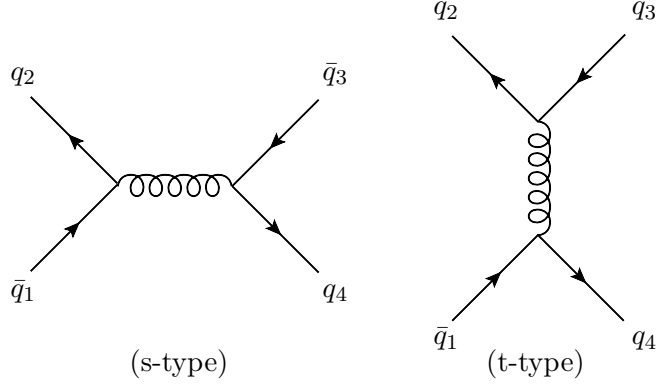}

\vspace{0.1cm} 

\begin{picture}(0,0)(0,0)
\put(-75,0){(s-type)}
\put(60,0){(t-type)}

\put(-120,28){$\bar q_1$}
\put(-120,110){$q_2$}
\put(-2,110){$\bar q_3$}
\put(-2,28){$q_4$}

\put(40,10){$\bar q_1$}
\put(40,135){$q_2$}
\put(115,135){$\bar q_3$}
\put(115,10){$q_4$}
\end{picture}

\caption{Amplitudes of $s$-type  (left) and of $t$-type (right),
for the partonic process $0 \to W^+W^- \bar{q}_1 q_2 \bar{q}_3
  q_4$. The $W$-bosons are not shown.
} \label{fig:st_ampl}
\end{center}
\end{figure}

We now consider the case of amplitudes involving two $\bar{q}q$ pairs
and the $W^+W^-$ pair.  We first discuss the color and flavor
structure of the tree-level amplitude $0 \rightarrow (\bar{q}_1 q_2
\bar{q}_3 q_4) + (W^{+} \rightarrow \nu_{\mu} + \mu^+) + (W^-
\rightarrow e^- + \bar{\nu}_{e})$ treating all particles as being in
the final state.  This process is described by Feynman diagrams with
two continuous fermion lines connected by a gluon exchange, with
$W$-bosons being emitted from either of the two quark lines.
Depending on the quark flavors {\it and} on the way the $W$-boson emissions
occur, we may have to assign quark fields in two different ways to the
fermion lines: $[\bar q_1 q_2]$, $[\bar q_3 q_4]$ and $[\bar q_1
q_4]$, $[\bar q_3 q_2]$. We refer to the first assignment as the
``$s$-channel amplitude'' and to the second assignment as the
``$t$-channel amplitude'', see Fig.~\ref{fig:st_ampl}.

\begin{figure}[t]
 \begin{center}
  \includegraphics[height=6cm]{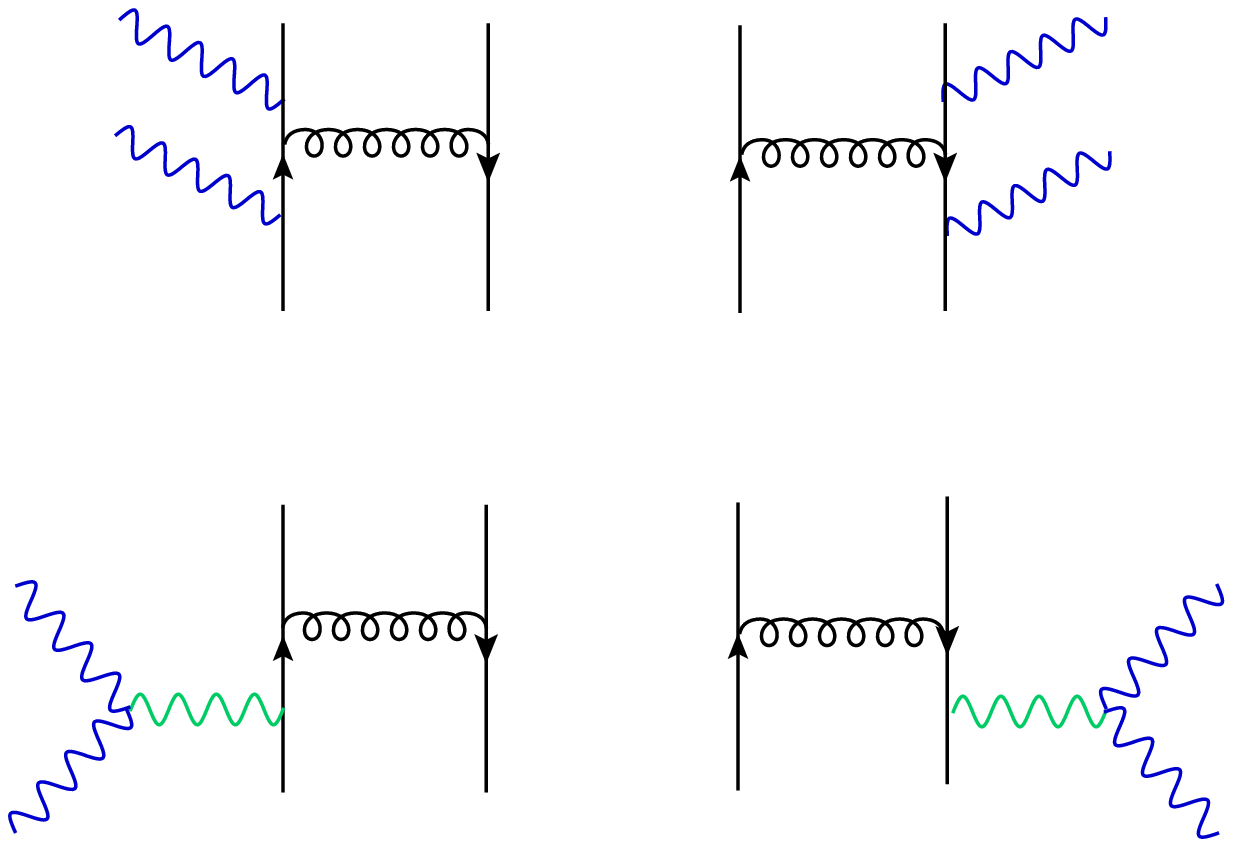}

\begin{picture}(0,0)(0,0)
\put(-72,113){$\bar u$}
\put(-28,113){$c$}

\put(-72,187){$u$}
\put(-28,187){$\bar c$}

\put(-72,15){$\bar u$}
\put(-28,15){$c$}

\put(-72,89){$u$}
\put(-28,89){$\bar c$}

\put(-123,155){$W^+$}
\put(-123,180){$W^-$}

\put(-95,27){$\gamma/Z$}
\put(-145,65){$W^-$}
\put(-145,12){$W^+$}

%------------------------------

\put(65,113){$c$}
\put(21,113){$\bar u$}

\put(65,187){$\bar c$}
\put(21,187){$u$}

\put(106,155){$W^-$}
\put(106,180){$W^+$}

\put(65,15){$c$}
\put(21,15){$\bar u$}

\put(65,89){$\bar c$}
\put(21,89){$u$}

\put(75,27){$\gamma/Z$}
\put(125,65){$W^+$}
\put(125,12){$W^-$}

\end{picture}

  \caption{Sample tree-level diagrams for $B_0(\bar{u},u,\bar{c},c$). When
    both $W$-bosons couple directly to the quark line, the flavors of
    the quarks determine the ordering of the $W$-bosons.} \label{fig:qqwwqqtree1}
 \end{center}
\end{figure}

\begin{figure}[t]
 \begin{center}
  \includegraphics[height=2cm]{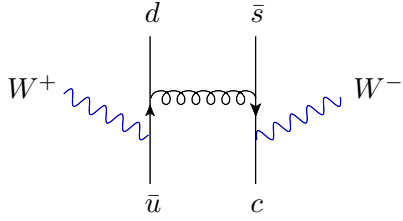}

\begin{picture}(0,0)(0,0)

\put(-74,48){$W^+$}
\put(-22,77){$d$}
\put(-22,5){$\bar u$}

\put(57,48){$W^-$}
\put(18,77){$\bar s$}
\put(18,5){$c$}

\end{picture}

\caption{Sample tree-level diagram for $B_0(\bar{u},d,\bar{s},c$).} 
\label{fig:qqwwqqtree2}
 \end{center}
\end{figure}

We begin by considering the $s$-channel tree-level amplitude.  In this
case, the color decomposition reads
\begin{equation} 
\begin{split}
\mathcal{B}^{\mathrm{tree}}(\bar{q}_1,q_2;\bar{q}_3,q_4;\nu_{\mu},\mu^+,e^-,\bar{\nu}_{e}) =&g_s^2 \biggl( \frac{g_W}{\sqrt{2}} \biggr)^4 P_{W}(s_{e^+\nu_e}) 
P_{W}(s_{\mu^{-} \nu_{\mu}}) 
\\
& 
\times 
\bigl(\delta_{\bar{i}_1 i_4} \delta_{\bar{i}_3 i_2} 
-  \frac{1}{N_c} \delta_{\bar{i}_1 i_2} \delta_{\bar{i}_3 i_4} \bigr) 
B_{0}(\bar{q}_1,q_2; \bar{q}_3,q_4).
\end{split} \label{qqtree}
\end{equation}
We can further split the $B_0$ amplitude into two separate types.  The
amplitude of the first type appears if a quark line radiates both $W$-bosons and the other quark line radiates none. The $W$-boson can be
radiated either directly from the quark line, or through an exchange of
an intermediate $\gamma/Z$.  The amplitude of the second type arises
when one $W$-boson is radiated from each of the quark lines.  Examples
of the corresponding contributions are shown in
Figs.~\ref{fig:qqwwqqtree1},\ref{fig:qqwwqqtree2} for specific flavor
assignments; it is clear how this classification generalizes to other
flavors.

We begin by discussing amplitudes of the first type.  Since we set the
CKM matrix equal to the identity matrix, flavors of fermions can not
change along the fermion lines, so that flavors of $\bar{q}_1$ and
$q_2$ as well as of $\bar{q}_3$ and $q_4$ are equal.  Thus, for a set
of flavor assignments, for which this contribution is allowed, there
are four diagrams that contribute to the amplitude $B_0$. Examples are
shown in Fig. \ref{fig:qqwwqqtree1}.  We write the color-ordered
amplitude as
\begin{equation} \label{uucc}
\begin{split}
& B_0(\bar{q}_1, q_2; \bar{q}_3, q_4) = B_0^{(WW)} \bigl( [\bar{q}_1,W,W, q_2], [\bar{q}_3,q_4] \bigr)  
+B_0^{(WW)} \bigl( [\bar{q}_1, q_2], [\bar{q}_3, W,W,q_4] \bigr) \\
& \;\;\;\;\;\;\;\;\;
+C^{(q_2,h_2)} B_0^{(\gamma/Z)} \bigl( [\bar{q}_1,\gamma/Z, q_2], [\bar{q}_3, q_4] \bigr) 
 +C^{(q_4,h_4)} B_0^{(\gamma/Z)} \bigl( [\bar{q}_1, q_2],[\bar{q}_3, \gamma/Z, q_4] \bigr),
\end{split} 
\end{equation}
where $h_{2,4} = \{-,+\}$ are the helicities of quarks $q_2$ and
$q_4$.  We note that since $W$-bosons couple only to left-handed
quarks, the first term in Eq.(\ref{uucc}) is zero for $h_2 = 1$, and
the second term is zero for $h_4 = 1$. The factors $C^{(q,h)}$ are
given in Eq.~\eqref{gamZ}.  The second term in Eq.~(\ref{uucc}) can be
obtained from the first term by swapping momenta $p_{\bar{q}_1}
\leftrightarrow p_{\bar{q}_3}$ and $p_{q_2} \leftrightarrow
p_{q_4}$. The same swap can be used to obtain the fourth term in
Eq.~(\ref{uucc}) from the third.

We turn to the discussion of the amplitudes of the second type, which
correspond to the emission of the $W^+$ boson off one quark line and
the $W^-$ boson off the other quark line. As a result of the emission,
flavors change along each fermion line.  An example of a diagram
contributing to this amplitude is shown in Fig.~\ref{fig:qqwwqqtree2}.
As there is no contribution of the neutral vector boson in this case,
the amplitude is nonzero only for $h_2 = h_4 = -1$.  The choice of
flavors for $\bar q_1, \bar q_3$ determine which $W$-boson is radiated
from which quark line.

According to the flavors of the four quarks, only one of the $s$- or
$t$-channel amplitudes can contribute, or both.  Since the $t$-channel
amplitude is obtained by replacing $q_2 \leftrightarrow q_4$ in the
$s$-channel amplitude, everything that has been said about the latter
applies to the former. Note that the replacement $q_2 \leftrightarrow
q_4$ also involves color indices, so that non-trivial
color-correlations appear in the interference of $s$- and $t$-channel
amplitudes when both are allowed by flavor.

For the computation of real emission corrections we need four-quark
amplitudes with additional gluon in the final state $0 \rightarrow
(\bar{q}_1 q_2 \bar{q}_3 q_4) + (W^{+} \rightarrow \nu_{\mu} + \mu^+)
+ (W^- \rightarrow e^- + \bar{\nu}_{e}) + g$. It is clear that the
presence of an additional gluon does not modify the separation of
amplitudes into $s$- and $t$-channel amplitudes, so that much of what
has been said about the tree-level amplitudes remains applicable.  In
particular, the flavor structure is identical to the tree-level case
discussed above.  On the other hand, the color decomposition
differs. For instance, for the $s$-channel amplitude, it reads
\begin{equation}
 \begin{split}
& \mathcal{B}^{\mathrm{tree}}(\bar{q}_1,q_2,\bar{q}_3,q_4,g;\nu_{\mu},\mu^+,e^-,\bar{\nu}_{e}) = g_s^3 \biggl( \frac{g_W}{\sqrt{2}} \biggr)^4 
P_{W}(s_{\nu_{\mu}\mu^+}) P_{W}(s_{e^{-} \bar{\nu}_{e}})  \\
& \;\;\;\;\;\;\;\;
\times \biggl[ \delta_{\bar{i}_3 i_2} T^a_{\bar{i}_1i_4} B_0(\bar{q}_1,q_2,\bar{q}_3,q_4,g) + \frac{1}{N_c} \delta_{\bar{i}_3 i_4} T^a_{\bar{i}_1 i_2} B_0(\bar{q}_1,g,q_2,\bar{q}_3,q_4) \\
&\;\;\;\;\;\;\;\;\;+ \delta_{\bar{i}_1 i_4} T^a_{\bar{i}+3 i_2} 
 B_0(\bar{q}_1,q_2,g,\bar{q}_3,q_4) 
+  \frac{1}{N_c}\delta_{\bar{i}_1 i_2} T^a_{\bar{i}_3 i_4} B_0(\bar{q}_1,q_2,\bar{q}_3,g,q_4) \biggr].
 \end{split}
\end{equation}

Similar considerations apply to virtual corrections but the color
decomposition is more involved in this case.  For the $s$-channel
virtual QCD amplitude it reads
 \begin{equation}
   \begin{split}
& \mathcal{B}^{\mathrm{1L}}(\bar{q}_1,q_2,\bar{q}_3,q_4;\nu_{\mu},\mu^+,e^-,\bar{\nu}_{e}) = 
g_s^4 \biggl( \frac{g_W}{\sqrt{2}} \biggr)^4 
P_{W}(s_{\nu_{\mu},\mu^+}) P_{W}(s_{e^{-} \bar{\nu}_{e}})  \\
   & \;\;\;\;\;\;\;
\times \bigl( \delta_{\bar{i}_1 i_4} \delta_{\bar{i}_3 i_2}
B_1^{(1)}(\bar{q}_1,q_2,\bar{q}_3,q_4) + \delta_{\bar{i}_1 i_2}
\delta_{\bar{i}_3 i_4} B_1^{(2)}(\bar{q}_1,q_2,\bar{q}_3,q_4)
\bigr)\,. 
   \end{split} \label{qqvirt}
 \end{equation}

\begin{figure}[t]
\begin{center}
\includegraphics[height=3cm]{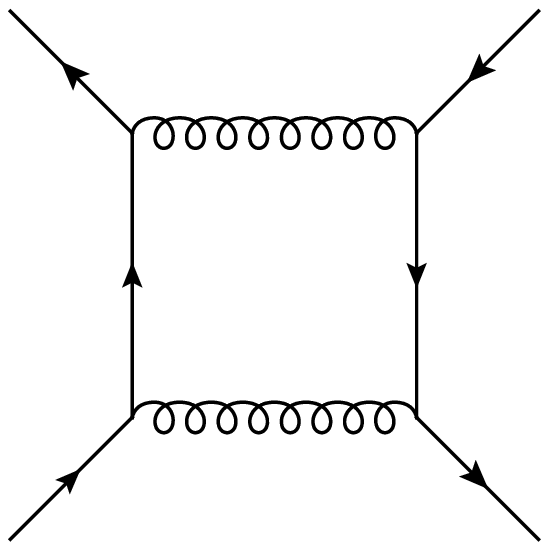}
\hspace*{0.4cm} 
\includegraphics[height=3cm]{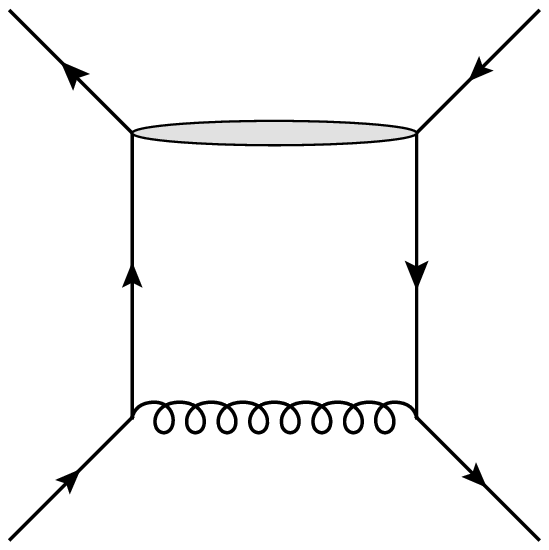}
\hspace*{0.4cm}
\includegraphics[height=3cm]{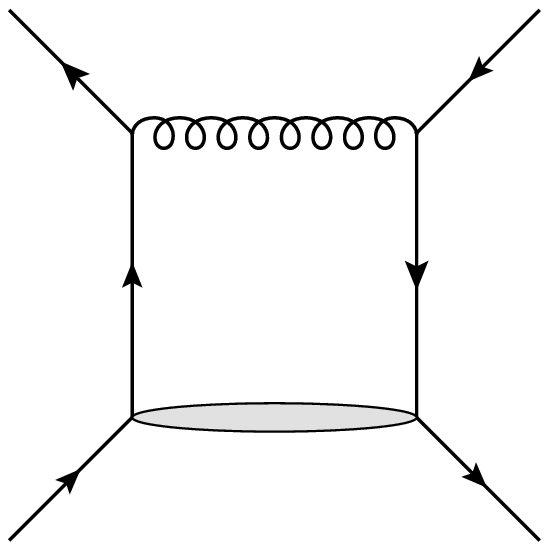}
\hspace*{0.4cm}
\includegraphics[height=3cm]{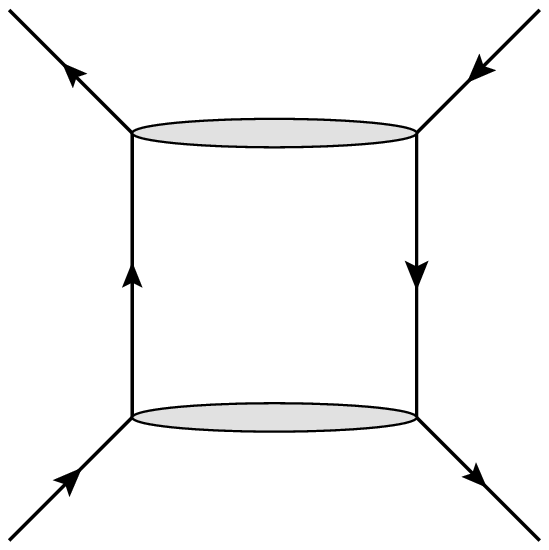}

\begin{picture}(0,0)(0,0)
\put(-205,5){$\bar q_1$}
\put(-205,105){$q_2$}
\put(-115,105){$\bar q_3$}
\put(-115,5){$q_4$}
\end{picture}

\caption{Parent diagrams for one-loop primitive amplitudes $B_1^{(a,b,c,d)}$
for $0 \to (\bar{q}_1 q_2 \bar{q}_3 q_4) + W^+W^-$, where the flavors of 
the quarks are not specified. The $W$-bosons are not shown. Shaded 
areas represent dummy lines 
which are not cut.} \label{fig:qqvirtampl}
\label{fig0a}
\end{center}
\end{figure}

The amplitudes in Eq.(\ref{qqvirt}) are written through primitive 
amplitudes as  
\begin{equation}
 \begin{split}
&  B_1^{(1)}(\bar{q}_1,q_2,\bar{q}_3,q_4) = \bigl( N_c - \frac{2}{N_c} \bigr) B_1^{(a)}(\bar{q}_1,q_2,\bar{q}_3,q_4) -\frac{2}{N_c} B_1^{(a)}(\bar{q}_1,q_2,\bar{q}_3,q_4) \\
& -\frac{1}{N_c} B_1^{(b)} (\bar{q}_1,q_2,\bar{q}_3,q_4)-\frac{1}{N_c}B_1^{(c)} (\bar{q}_1,q_2,\bar{q}_3,q_4) +n_f B_1^{(d)}(\bar{q}_1,q_2,\bar{q}_3,q_4),
 \end{split} \label{qqprim1}
\end{equation}
and
\begin{equation}
 \begin{split}
&  B_1^{(2)}(\bar{q}_1,q_2,\bar{q}_3,q_4) = \frac{1}{N_c^2} B_1^{(a)}(\bar{q}_1,q_2,\bar{q}_3,q_4) + \bigl( 1+\frac{1}{N_c^2} \bigr) B_1^{(a)}(\bar{q}_1,q_2,\bar{q}_3,q_4) \\
& +\frac{1}{N_c^2} B_1^{(b)} (\bar{q}_1,q_2,\bar{q}_3,q_4)+\frac{1}{N_c^2}B_1^{(c)} (\bar{q}_1,q_2,\bar{q}_3,q_4) - \frac{n_f}{N_c} B_1^{(d)}(\bar{q}_1,q_2,\bar{q}_3,q_4).
 \end{split} \label{qqprim2}
\end{equation}
Parent diagrams for the primitive amplitudes $B_1^{(a,b,c,d)}$ are
shown in Fig.~\ref{fig:qqvirtampl}.  The only primitive amplitude that
receives contributions from six-point one-loop diagrams is $B^{(a)}$.
Primitives $B^{(b)}$ and $B^{(c)}$ are simply the Born amplitudes
dressed by a gluon loop on one of the quark lines, while $B^{(d)}$
corresponds to a fermion loop contribution.  For convenience, we again
use dummy lines in Fig.~\ref{fig:qqvirtampl}; they allow us to
consider every primitive amplitude as having a parent diagram with
(formally) six propagators. We recall that $W$-bosons couple to dummy
lines, but that these lines cannot be cut. Berends-Giele recursion
relations are modified in these cases to ensure that the correct
primitive amplitudes are recovered.

\subsection{Checks on the calculation}

Various checks were carried out at all stages of the calculation. The
squared matrix elements for the leading order and real emission
processes were checked against {\sf MadGraph} \cite{Alwall:2007st} for
a few phase space points. This was done for all flavor combinations
and all initial state parton configurations. Gauge invariance of
various amplitudes was checked for both the external gluons and the
$W$-bosons (artificially setting the masses of the latter to zero), at
leading and next-to-leading order.  The subtraction terms of the
Catani-Seymour dipole method were checked to cancel with the real
emission terms in the limit when emitted partons become soft and/or
collinear.  We checked the double and single infrared poles of the
virtual contribution, both at the level of primitive amplitudes and at
the level of virtual matrix elements squared.  These terms were also
checked to cancel with the integrated dipoles.  We also checked the
independence of the cross-section of the $\alpha$-parameter
\cite{Nagy:1998bb,Nagy:2003tz}. Finally, the full one-loop amplitude
is checked against an OPP-based, but otherwise completely independent
diagrammatic computation, at a few phase space points.  We note that
over six hundred Feynman diagrams are involved in a such a
calculation.

By default, our calculation was performed in double precision.  For
each phase space point, the double and single poles were checked
against the analytically known results, and the coefficients of the
OPP expansion were checked to have accurately solved the system of
linear equations. If either of those checks failed, the amplitude at
that phase space point was recalculated using quadruple precision. We
found that around $0.4\%$ of primitive amplitudes had to be
recalculated this way.

\section{Phenomenology}
\label{pheno}

In this Section, we discuss phenomenological aspects of $W^+W^-jj$
production at the Tevatron and the LHC.  At the Tevatron, this process
is a background to Higgs-boson production in association with two
jets. We employ set of cuts discussed in the context of the
Higgs-boson search in Ref.~\cite{higgscdf} and study related
phenomenology.  At the LHC, we consider the collision energy of
$7~{\rm TeV}$ and we show that the number of dilepton events related
to $W^+W^-jj$ production is sufficiently large to study this process
in detail.

Before moving on to a dedicated discussion, we briefly describe
general features of our computation.  The $W$-bosons are always
produced on mass-shell and decay leptonically $W^+ W^- \to \nu_{\mu}
\mu^+ e^- \bar{\nu}_e$. We note that, neglecting non-resonant
contributions, the results for all lepton flavors $ l^+l^- = \{
e^+e^-,e^+ \mu^-,\mu^+ e^-,\mu^+ \mu^- \}$ can be obtained by
multiplying our results by four.  

The mass and width of the $W$-boson are taken to be $M_W = 80.419~{\rm
  GeV}$ and $\Gamma_W = 2.141$~{\rm GeV}, respectively. 
The width of
the $Z$-boson is taken to be $\Gamma_Z = 2.49~\mathrm{GeV}$. The
propagators for these particles take the Breit-Wigner form.  
The
electroweak gauge couplings are computed using $\alpha_{\rm QED} (M_Z)
= 1 / 128.802$ and $\sin^2 \theta_W = 0.2222$. We use MSTW08LO parton
distribution functions for leading order and MSTW08NLO for
next-to-leading order computations \cite{Martin:2009iq}.  The strong
coupling constant $\alpha_s(M_Z)$ is part of the MSTW fit. It equals
to $0.13939$ ($0.12018$) at leading- and next-to-leading order,
respectively.

\subsection{Results for the Tevatron}

\begin{figure}[t]
\begin{center}
\includegraphics[width=7.5cm]{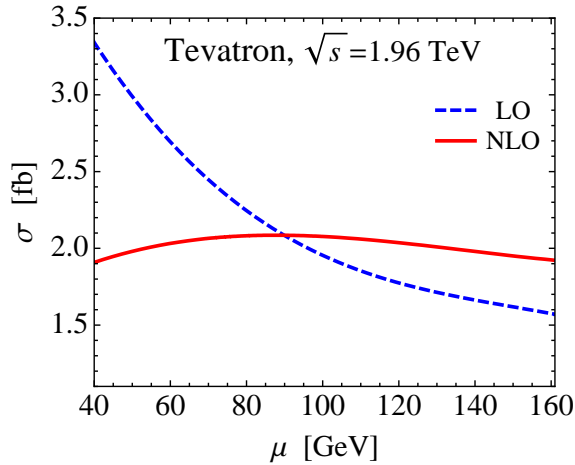}
\caption{The dependence on renormalization and factorization scale of
  the cross section for $p\bar{p}\rightarrow \nu_\mu\mu^+e^-
  \bar{\nu}_e\,j j$ at $\sqrt{s}=1.96$ TeV, where
  $\mu=\mu_R=\mu_F$. Predictions at both LO and NLO in QCD are shown. }
 \label{teva1}
\end{center}
\end{figure}

By the end of Run II, the Tevatron will have collected just over 10
fb$^{-1}$ of data for use in the search for the Higgs boson.  At the
very least, the two Tevatron experiments will be able to improve upon
the exclusion limits for the Higgs bosons presented earlier in
Ref.~\cite{:2009je}.  The search strategy is to separate relevant
processes, depending on the number of jets produced with the Higgs
boson. As follows from the analysis in Ref.~\cite{Campbell:2010cz},
ten percent of all events with the Higgs boson at the Tevatron contain
two or more jets.  The process $pp \to W^+W^-jj$ is a SM
background of significant importance.  In Ref. \cite{Campbell:2010cz},
the NLO QCD cross section for Higgs $+2~{\rm jet}$ production with the
decay $H\rightarrow
W^-(\rightarrow\mu^-\bar{\nu}_\mu)W^+(\rightarrow\nu_e e^+)$ is
calculated.  For the Higgs-boson mass of $160~{\rm GeV}$, the
cross-section value $\sigma_{\rm NLO} = 0.2~{\rm fb}$ is found (we do
not show the uncertainties which are significant).  This cross-section
is obtained with cuts that are similar to those used by the CDF
collaboration in their Higgs-boson search
\cite{higgscdf}. Specifically, jets are defined using the
$k_\perp$-algorithm, with $\Delta R_{j_1j_2}>0.4$.  Jets must have
$p_{\perp,j} > 15$ GeV and must be in the central region of the
detector, $|\eta_j| < 2.5$.  It is required that two leptons, one with
transverse momentum $p_{\perp,l_1}\! >\! 20$ GeV and rapidity
$|\eta_{l_1}|\! <\! 0.8$ and the other with transverse momentum
$p_{\perp,l_2}\! >\! 10$ GeV and rapidity $|\eta_{l_2}| \!<\!
1.1$, appear in the event.  The invariant mass of the lepton pair is
required to be larger than $m_{l_1l_2}>16$ GeV.  Both leptons must be
isolated. The specific requirement to this effect is that any jet
within $\Delta R = 0.4$ of a lepton must have a transverse momentum
which is smaller than $0.1~p_{\perp,l}$.  The CDF collaboration uses a
particular constraint on the missing transverse momentum.  They
introduce a function
\begin{eqnarray}
\slashed{E}^{\text{spec}}_{\perp} = \slashed{E}_\perp \sin{[\text{min}(\Delta\phi,\frac{\pi}{2})]}, \nonumber
\end{eqnarray}
with $\Delta\phi$ being the angle between the missing transverse
momentum vector $\slashed{E}_\perp$ and the nearest lepton or jet.  An
event is accepted if $\slashed{E}^{\text{spec}}_{\perp } > 25$ GeV.

We present NLO QCD results for the process $p\bar{p}\rightarrow
W^+(\rightarrow\nu_\mu\mu^+) W^-(\rightarrow e^- \bar{\nu}_e) j j$, at
$\sqrt{s}=1.96$ TeV, using the kinematic cuts that we just described.
This allows us to study this process as a 
background to the Higgs-boson production.  In Fig. \ref{teva1} we show
the scale dependence of the cross section for the process
$p\bar{p}\rightarrow W^+(\rightarrow\nu_\mu\mu^+)\;W^-(\rightarrow e^-
\bar{\nu}_e) j j$, both at LO and NLO in
perturbative QCD (pQCD), with the scale ranging between $M_W/2$ and
$2M_W$. 

The leading order cross-section is $\sigma_{\rm LO} = 2.5 \pm 0.9$
fb.  This result is interesting since {\it its uncertainty alone
  exceeds the cross-section for the production of the Higgs boson in
  association with two jets by about a factor between four and five.}
Clearly, there is no way to discuss observation of the Higgs boson in
this channel unless the theoretical uncertainty on $W^+W^-jj$ is
improved.  The situation indeed improves once NLO QCD corrections are
computed.  We find $\sigma_{\rm NLO} = 2.0 \pm 0.1$ fb -- a
significant reduction in scale uncertainty. However, even after that
reduction, we find that the {\it uncertainty} on the $W^+W^-jj$
production cross-section is very much comparable to the {\it absolute
  value} of the Higgs-boson production cross-section in association
with two jets.  For this set of cuts, the NLO QCD computations lead to
a prediction of about 80 $e \mu^+ jj,e^+\mu jj,e^+e^- jj,\mu^+\mu^-
jj$ events during Run II, using the discussed set of cuts and assuming
$100\%$ efficiency.

\begin{figure}[t]
\begin{center}
\includegraphics[width=6.5cm]{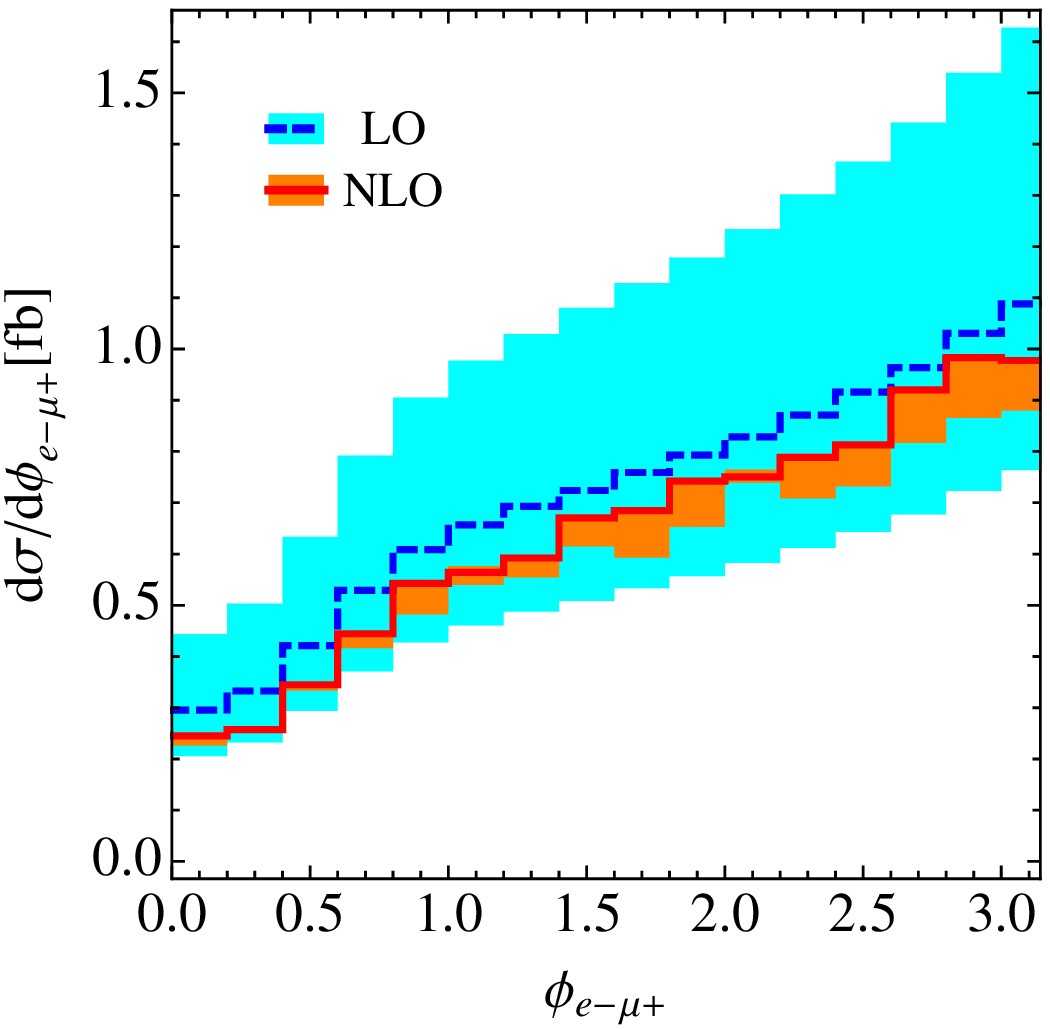} \hspace*{0.5cm} 
\includegraphics[width=6.5cm]{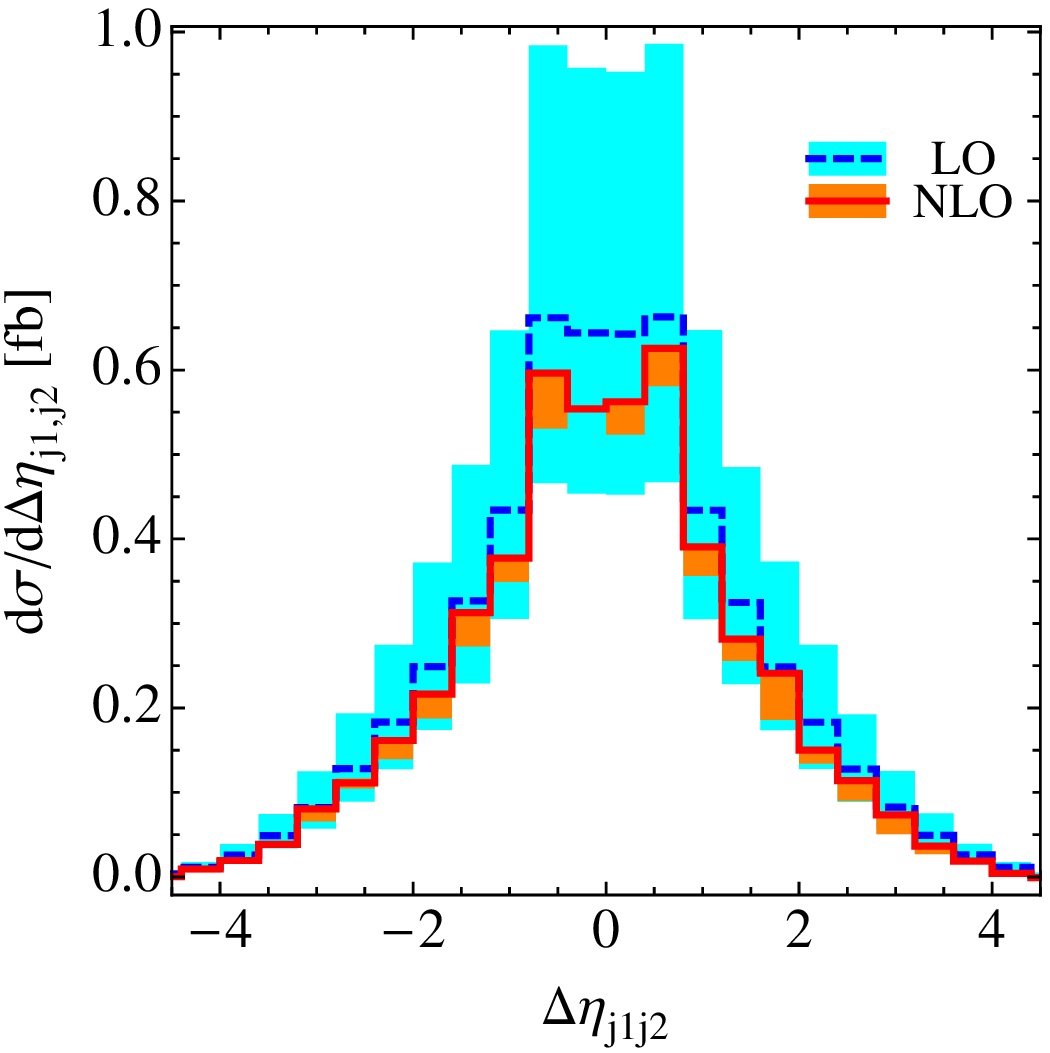}\\
\caption{Kinematic distributions showing the opening angle between the
  leptons, $\phi_{e^-\mu^+}$, and the difference in rapidity of the
  two hardest jets, for the process $p\bar{p}\rightarrow
  W^+(\rightarrow\nu_\mu\mu^+) W^-(\rightarrow e^- \bar{\nu}_e) j j$
  at the Tevatron running at $\sqrt{s} = 1.96 \, {\rm TeV}$. The bands
  show renormalization and factorization scale uncertainty for $M_W/2
  < \mu < 2M_W$, and the solid line is the prediction for $\mu=M_W$.}
\label{teva2}
\end{center}
\end{figure}

\begin{figure}[t]
\begin{center}
\includegraphics[width=6.5cm]{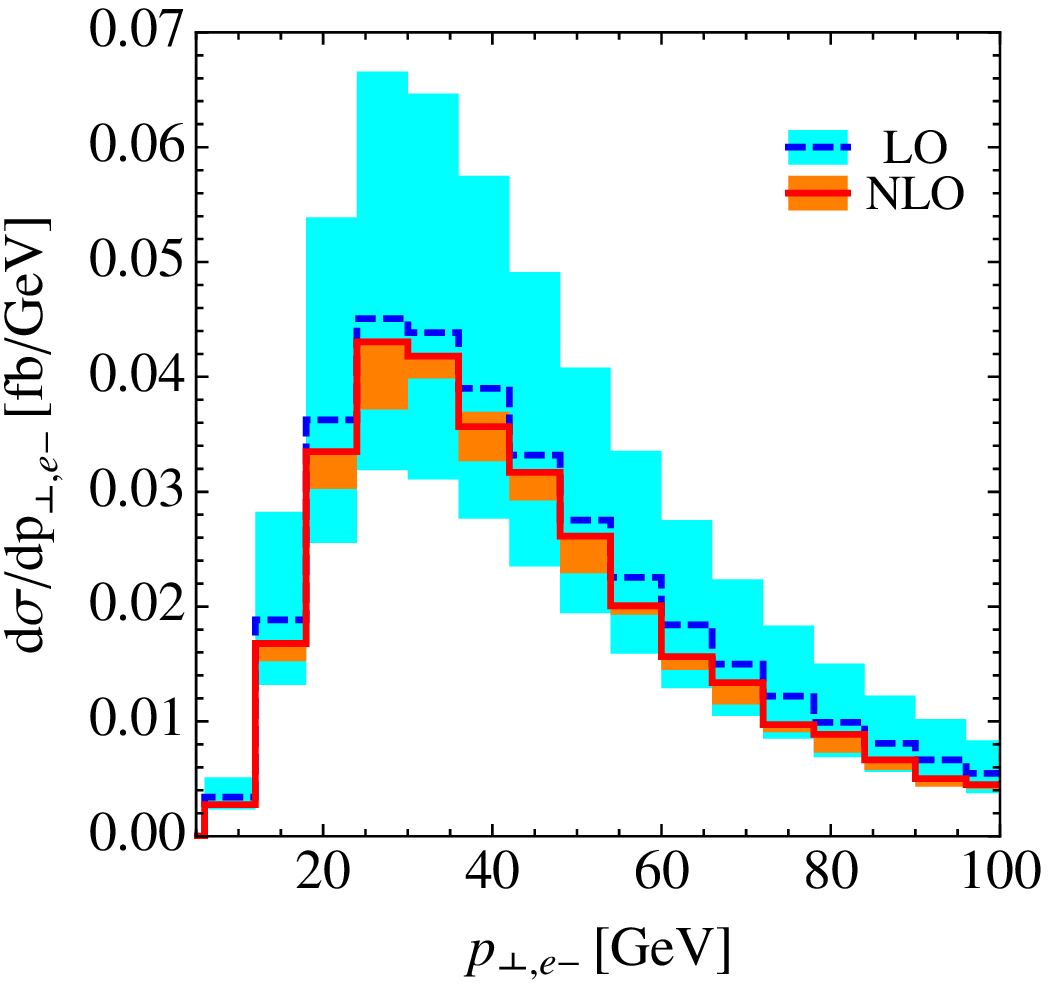} \hspace*{0.5cm} 
\includegraphics[width=6.5cm]{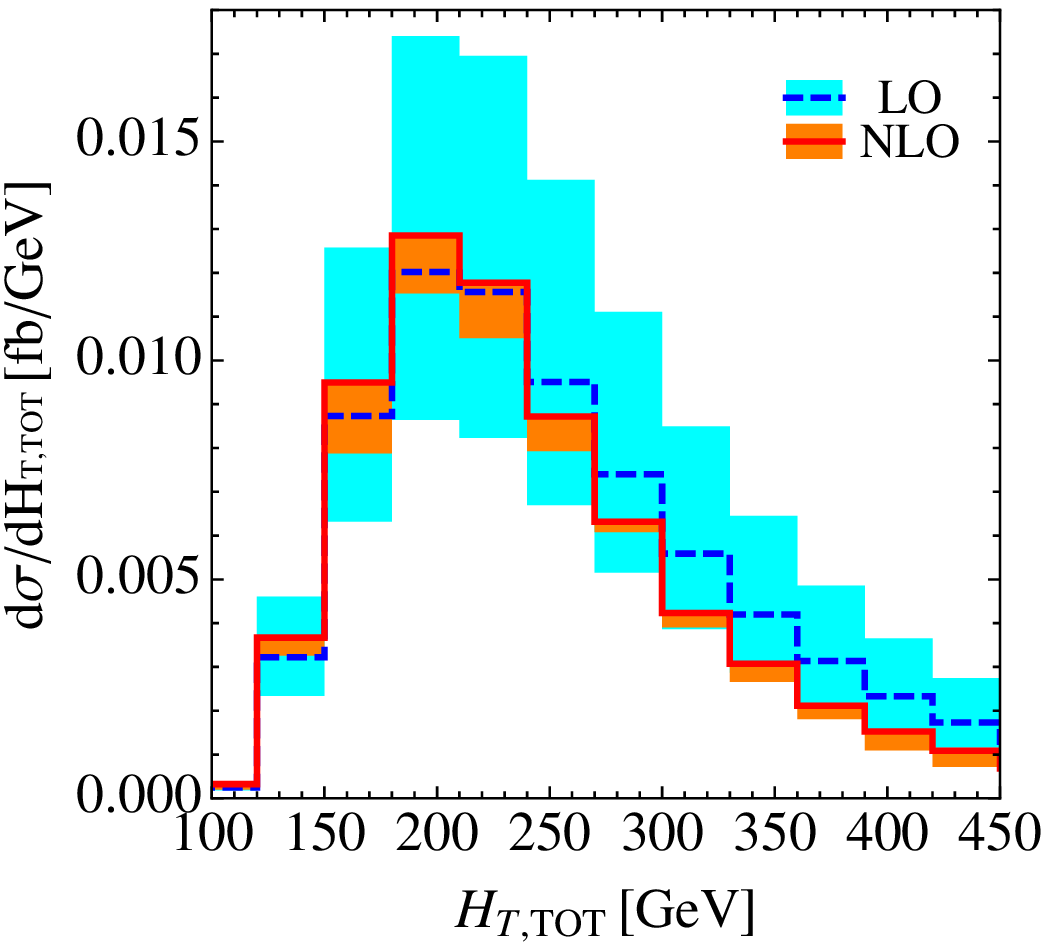}
\caption{Kinematic distributions showing the transverse momentum of a
  lepton and $H_{T,\text{TOT}}$, for the process $p\bar{p}\rightarrow
  W^+(\rightarrow\nu_\mu\mu^+) W^-(\rightarrow e^- \bar{\nu}_e) j j$
  at the Tevatron running at $\sqrt{s} = 1.96 \, {\rm TeV}$. The bands
  show renormalization and factorization scale uncertainty for $M_W/2
  < \mu < 2M_W$, and the solid line is the prediction for $\mu=M_W$.}
\label{teva3}
\end{center}
\end{figure}

There are other kinematic variables that one can use to improve upon
a discrimination between the Higgs-boson production and the $W^+W^-$
production.  For example, the opening angle of the two leptons is of
particular interest.  Indeed, if a pair of $W$-bosons is produced in
the decay of a scalar particle, their spins are anti-correlated. As a
result, leptons from their decay tend to have small relative angles in
the transverse plane. The $\phi_{e^-\mu^+}$ distribution in the case
of QCD $W^+W^-jj$ production is shown in Fig.~\ref{teva2} and the
leptons are seen to have a preference to be back-to-back, in strong
contrast to the Higgs-boson signal.  No noticeable shape changes occur
when the QCD corrections are included.  In the second pane of
Fig.~\ref{teva2}, we plot the rapidity difference between the two
hardest jets $\Delta \eta_{j_1 j_2} = \eta_{j_1} - \eta_{j_2}$, 
which is peaked at zero and falls off rapidly, with an
almost vanishing fraction of the cross section having a magnitude of
rapidity difference greater than four. Note that a  requirement
$|\Delta \eta_{j_1,j_2}|> 4$  is imposed when the Higgs boson 
is searched for in weak boson
fusion.

Finally, Fig.~\ref{teva3} shows the transverse momentum distribution
of the charged lepton and $H_{T,\text{TOT}}$ defined as the scalar sum of
the transverse momenta of all visible particles present in the final
state plus the missing transverse momentum, $H_{T,\rm TOT}= \sum_{j}
p_{\perp,j} + p_{\perp, \mu^+} + p_{\perp,e^-} + p_{\perp,
  \mathrm{miss}}$. It follows from Fig.~\ref{teva3} that 
the shape of lepton transverse momentum distribution does not change 
but the $H_{T,\text{TOT}}$ distribution becomes somewhat softer at NLO QCD.

\subsection{Results for the LHC}
\label{lhcresults}

 The LHC is set to run at 7 TeV until the end of 2012, collecting 2-5
 fb$^{-1}$ of data.  As a result, a non-negligible number of dilepton
 events, originating from $W^+W^-jj$, will be observed at the LHC
 during this and next year, which warrants a phenomenological study of
 this process.  The importance of $pp \to W^+W^- jj$ process as a
 background to Higgs-boson production has been discussed extensively
 in the literature (see e.g. \cite{Klamke:2007cu}), especially with
 reference to the weak boson fusion production mechanism, where
 designed cuts on the jets can dramatically boost the signal to
 background ratio. In this paper, we do not employ the weak boson
 fusion cuts, opting instead for a selection criteria that give sizable
 cross-sections for $pp \to W^+W^- jj$.  Our choice of cuts is
 inspired by those that are made in the first analyses of $t\bar{t}$
 production by ATLAS and CMS collaborations
 \cite{Aad:2010ey,Khachatryan:2010ez}.  We do, however, plot
 distributions which are interesting in the context of reducing the
 $W^+W^-jj$ background to the Higgs-boson searches in weak boson
 fusion. For example, we study the relative jet rapidity $\Delta
 \eta_{j_1 j_2} = \eta_{j_1}\!-\eta_{j_2}$ distribution and the
 opening azimuthal angle of the two leptons $\phi_{l_1 l_2}$.
 Given that the center-of-mass energy of collisions at the LHC after
 the longer shutdown at the end of 2012 is not fully decided yet, we
 also find it interesting to show the behaviour of the cross-section
 as a function of $\sqrt{s}$.
 \begin{figure}[t]
 \begin{center}
 \includegraphics[width=7cm]{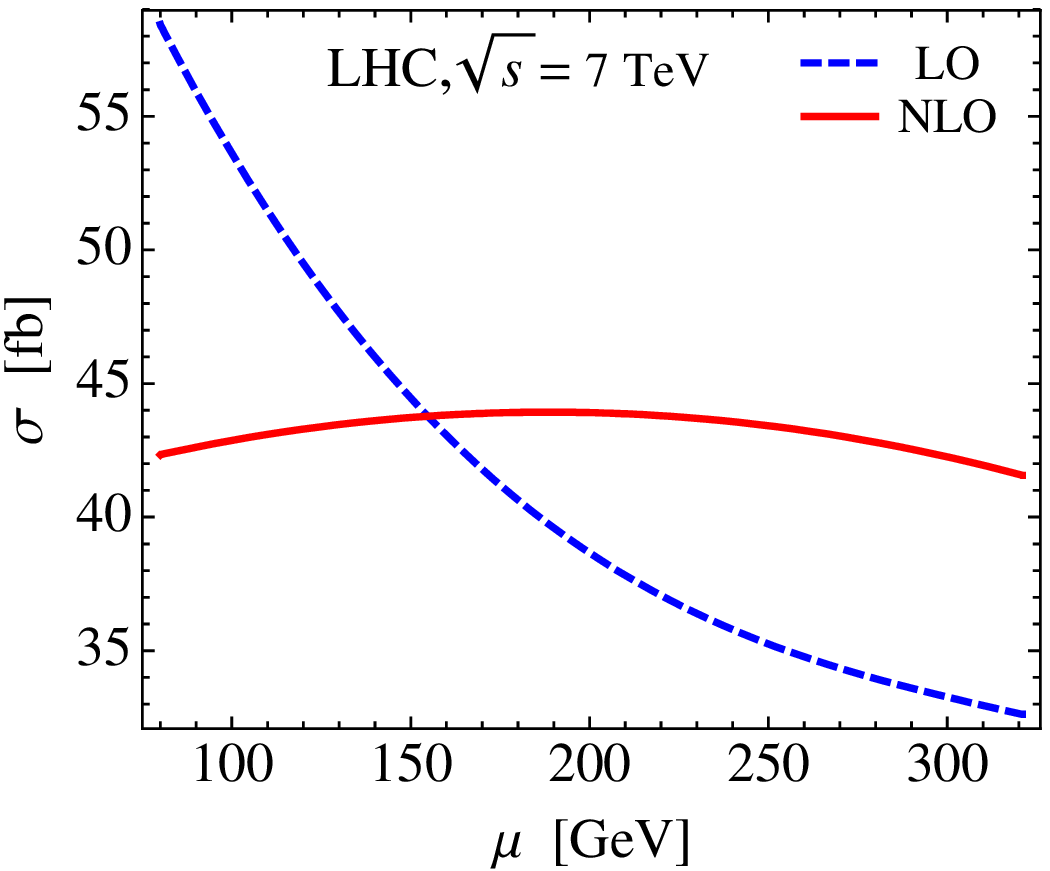}
 \qquad
\includegraphics[width=7cm]{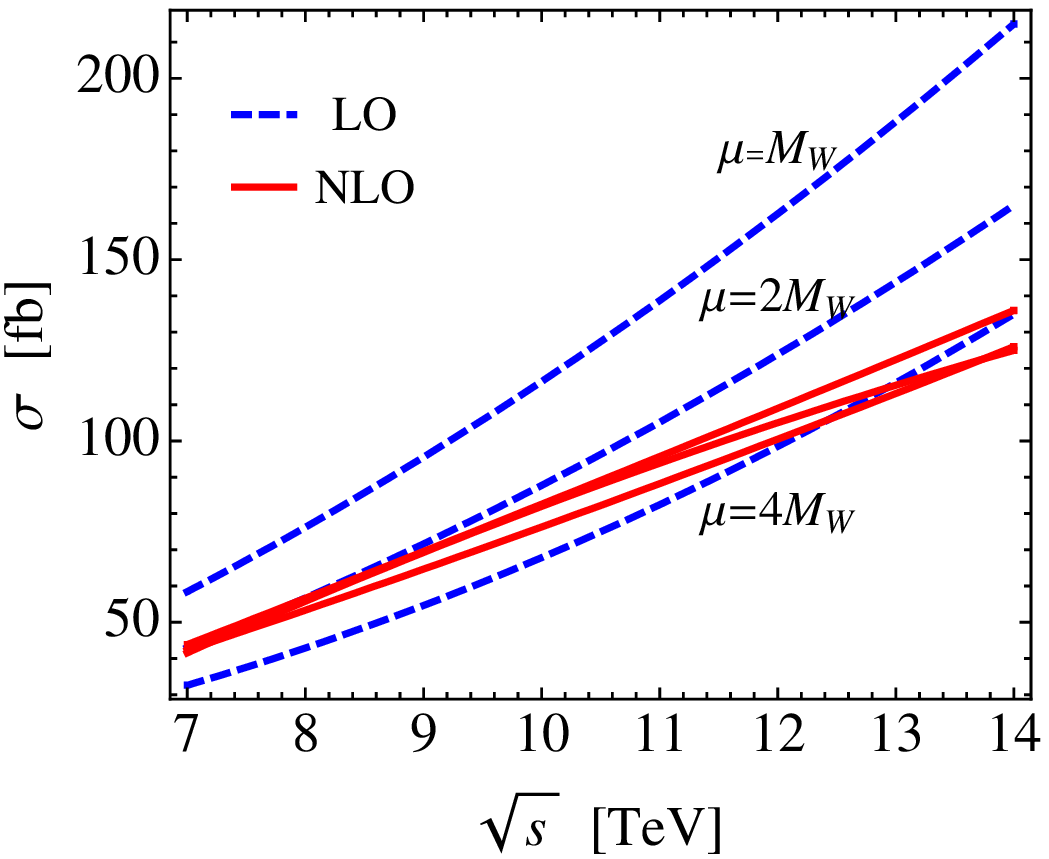} 
 
\caption{In the left pane, we show the production cross-section of the
  process $pp \to (W^+ \to {\nu}_{\mu} \mu^+)\, (W^- \to e^-
  \bar{\nu}_{e}) \, jj$ at the $7~{\rm TeV}$ run of the LHC in
  dependence of the factorization and renormalization scales $\mu_F =
  \mu_R = \mu$, at LO and NLO in perturbative QCD.  In the right pane,
  the dependence of the cross-section on the center-of-mass energy
  $\sqrt{s}$ is shown. LO results are shown in dashed blue; NLO
  results are in solid red. Three choices of $\mu$ are shown: $\mu =
  M_W,2M_W,4M_W$. }
 \label{figsigs}
 \end{center}
 \end{figure}

 We consider proton-proton scattering $ pp \to W^+ W^- j j$ at
 center-of-mass energy $\sqrt{s} = 7~\mathrm{TeV}$.  We impose the
 following cuts, inspired by $t \bar{t}$ searches at the LHC:
\begin{itemize}
 \item jets are defined 
using the anti-$k_{\perp}$ algorithm 
\cite{Cacciari:2008gp}
as implemented in FastJet~\cite{Cacciari:2005hq}, with 
\be
\Delta R_{j_1j_2}=\sqrt{(\eta_{j1} 
- \eta_{j2})^2 + (\phi_{j1} - \phi_{j2})^2} > 0.4;
\ee

\item jets are required to have transverse momentum 
$p_{\perp, j} > 30~{\rm GeV}$ and the rapidity $| \eta_j| < 3.2$;
 
\item charged leptons are required to have transverse momenta
$p_{\perp, l} > 20~{\rm GeV}$ and the rapidity  $| \eta_l| < 2.4$; 

\item missing transverse momentum is required to satisfy 
$p_{\perp, \rm miss} >30~{\rm GeV}$.

\end{itemize}

In the left pane of Fig.~\ref{figsigs} we show the dependence of the
cross-section $pp \to W^+ W^- \to \mu^+ \nu_{\mu} e \bar{\nu}_e \;jj$
at the $7~{\rm TeV}$ run of the LHC, on the factorization and
renormalization scales, which we set equal to each other. At
leading-order, the cross-section falls with the scale $\mu$, which is
attributable to the behaviour of the strong coupling $\alpha_s$.
Considering a range of factorization/renormalization scales $M_W < \mu
< 4 M_W$ and choosing the central value $\mu = 2 M_W$, we obtain a
cross-section $\sigma_{\mathrm{LO}} = 46 \pm 13~ \mathrm{fb}$.  At
next-to-leading order, the dependence on $\mu$ is dramatically reduced
and the cross-section becomes $\sigma_{\mathrm{NLO}} = 42 \pm 1 ~
\mathrm{fb}$.  Such a decrease in the scale dependence is typical of
NLO results, and indeed one of the primary motivations for performing
calculations at next-to-leading order in pQCD.  At the scale $\mu = 2
M_W$, the NLO corrections increase the cross-section by about
2\%. Assuming fifty percent efficiency, with $5~\mathrm{fb}^{-1}$ of
data at the $7~{\rm TeV}$ run of the LHC, we expect about 400 dilepton
events $e^+\mu^-, e \mu^+,e^+e^-,\mu^+\mu^-$.

\begin{figure}[t]
\begin{center}
\includegraphics[width=6.5cm]{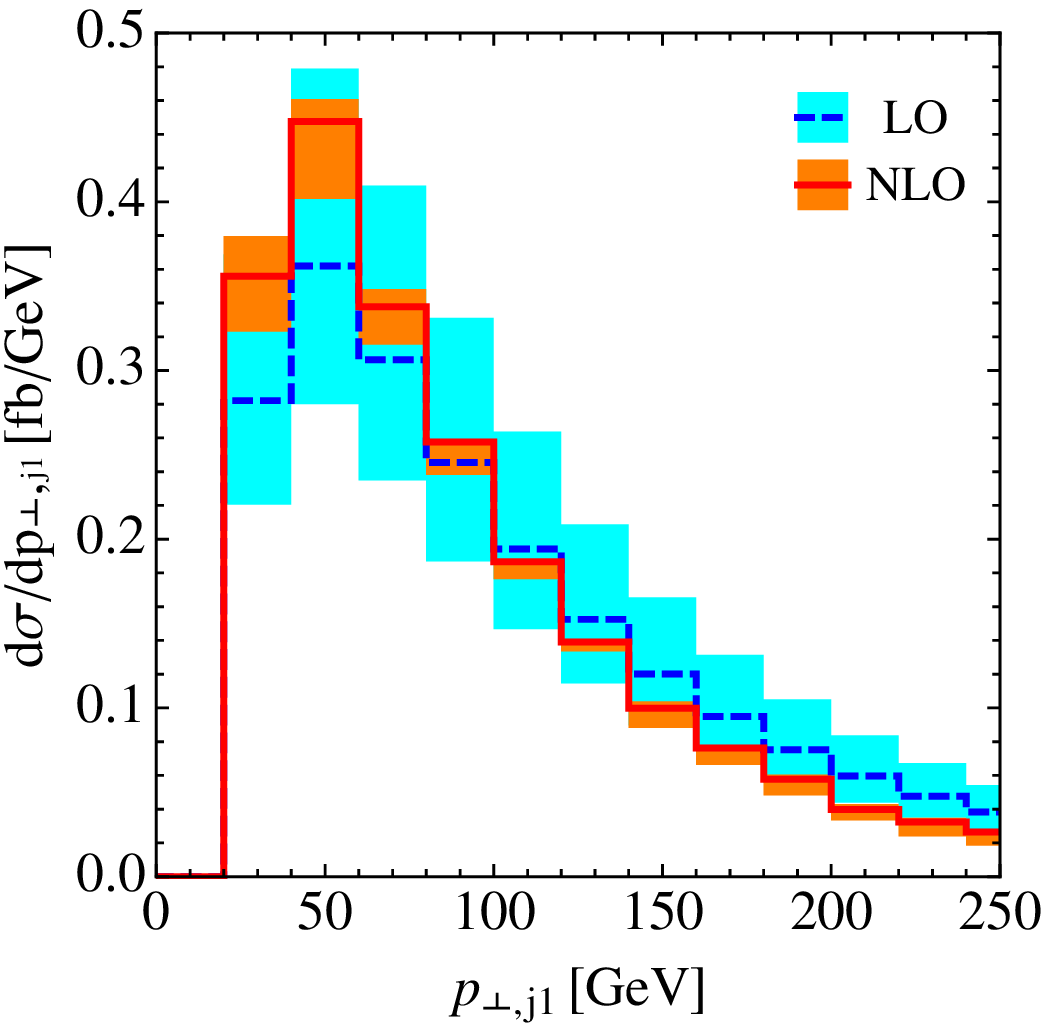}
\qquad
\includegraphics[width=6.5cm]{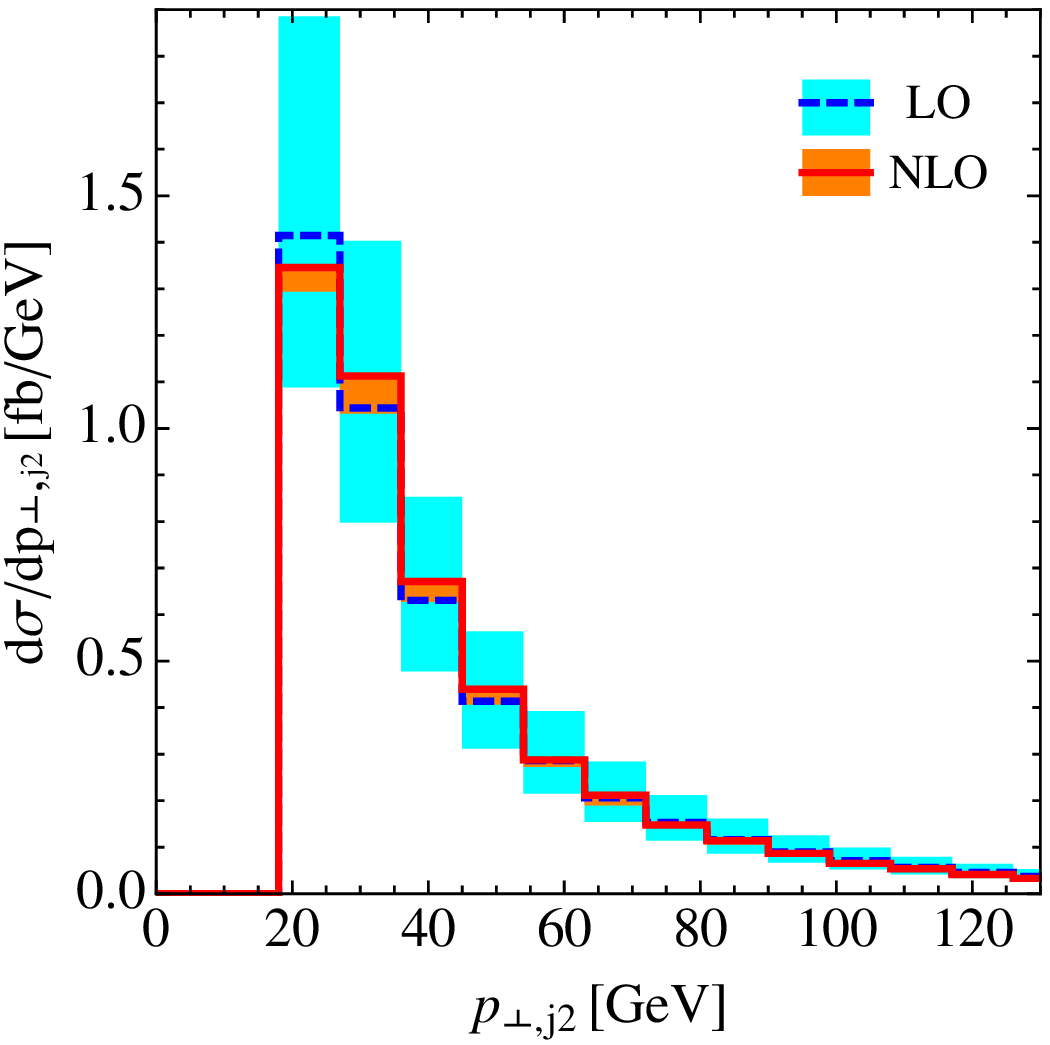}
\vspace{0.2cm}
\includegraphics[width=6.5cm]{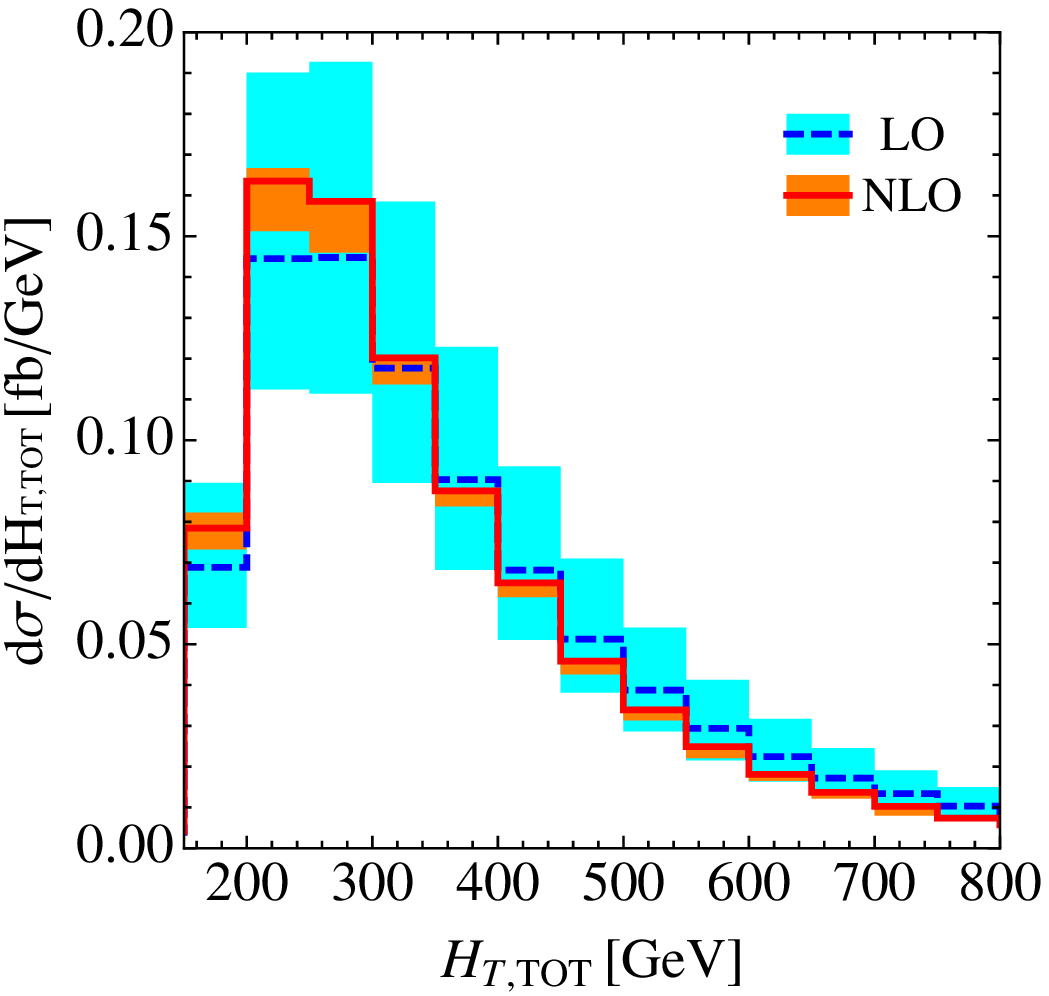}
\caption{Kinematic distributions for jets in the process $pp \to
  \nu_{\mu} \mu^+ e^- \bar{\nu}_e jj$ at the $7~$TeV run of the LHC at
  LO   and NLO in perturbative QCD. The bands show
  uncertainty on the renormalization and factorization scale $\mu$,
  for $M_W \leq \mu \leq 4M_W $, while the lines show results for $\mu
  = 2 M_W$. }
\label{figjets}
\end{center}
\end{figure}

It is interesting to know how the cross-section for $W^+W^-jj$
production changes with the collision energy.  In the right pane of
Fig.~\ref{figsigs}, we show that the dependence of the NLO
cross-section on the center-of-mass energy $\sqrt{s}$ is very close to
linear.  Again, the significant reduction in uncertainty in the NLO
prediction for the cross-section is obvious from
Fig.~\ref{figsigs}. It follows from Fig.~\ref{figsigs} that the
optimal\footnote{We define the ``optimal''
  renormalization/factorization scale as the value of $\mu$ for which
  next-to-leading order corrections are the smallest.}
renormalization/factorization scale increases with the center--of-mass
energy smoothly interpolating between $\mu = 2M_W$ at $7~{\rm TeV}$
and $\mu = 4~M_W$ at $14~{\rm TeV}$.

\begin{figure}[t]
\begin{center}
\includegraphics[width=6.5cm]{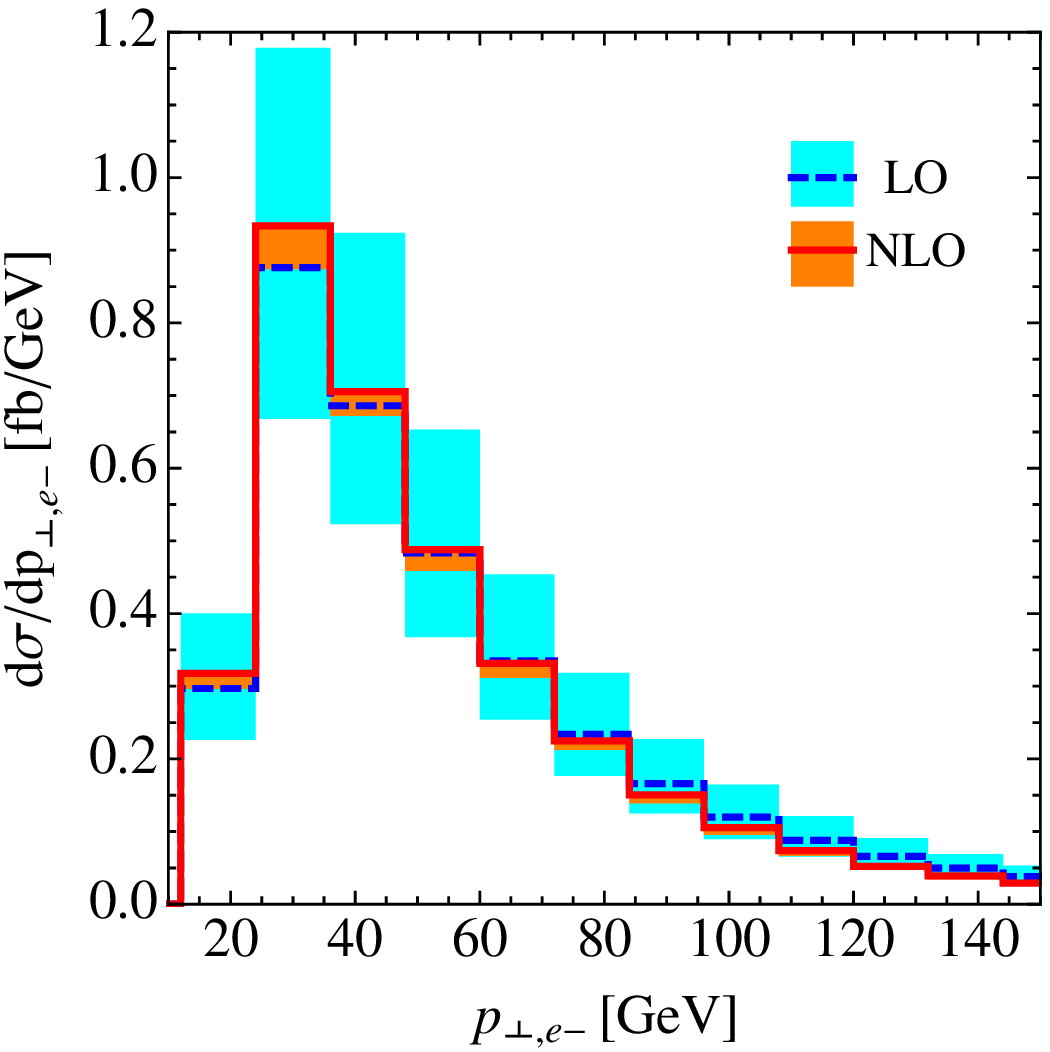}
\qquad
\includegraphics[width=6.5cm]{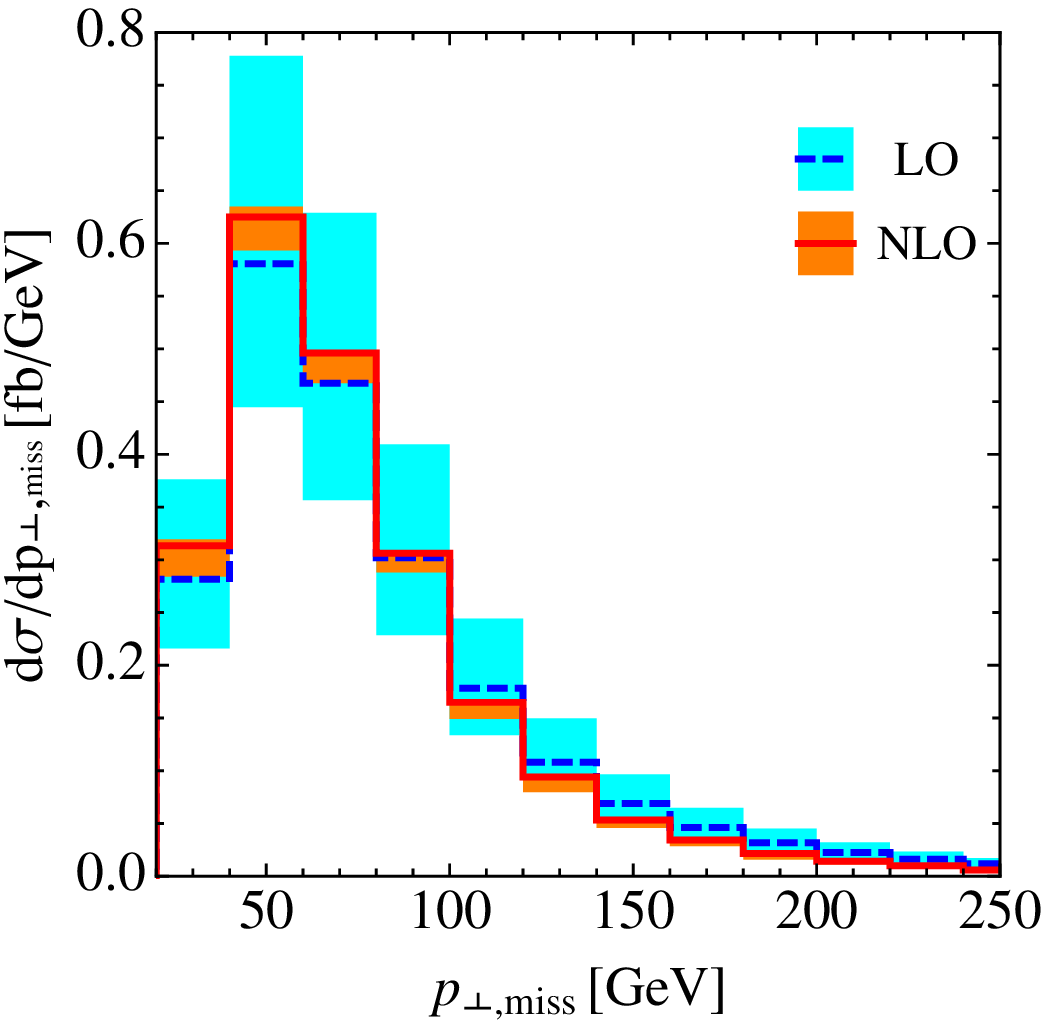}
\vspace{0.2cm}
\caption{Kinematic distributions for leptons in the process $pp \to
  \nu_{\mu} \mu^+ e^- \bar{\nu}_e jj$ at the $7~$TeV run of the LHC at
  LO and NLO in perturbative QCD.  The bands show
  uncertainty on the renormalization and factorization scale $\mu$,
  for $M_W \leq \mu \leq 4M_W$, while the lines show results for $\mu
  = 2 M_W $. }
\label{figleptons}
\end{center}
\end{figure}

We now turn to the discussion of kinematic distributions.  In
Fig.~\ref{figjets} we show the transverse momentum distribution of the
hardest and next-to-hardest jets and the distribution of the total
transverse energy $H_{T, \rm TOT}$. For all distributions, the scale
dependencies are reduced and shapes of the distributions are,
typically, not distorted. Note, however, that the NLO QCD corrections
make the jet transverse momenta distributions and the $H_{T,\rm TOT}$
distributions somewhat softer, which is caused, at least partially, by
our use of a constant, rather than a dynamic, renormalization scale in
the LO calculation.  We show lepton kinematic distributions in
Fig.~\ref{figleptons}.  Similar to jet distributions, lepton
transverse momentum and the missing energy distributions are softened
by the NLO QCD corrections.

\begin{figure}[t]
\begin{center}
\includegraphics[width=6.5cm]{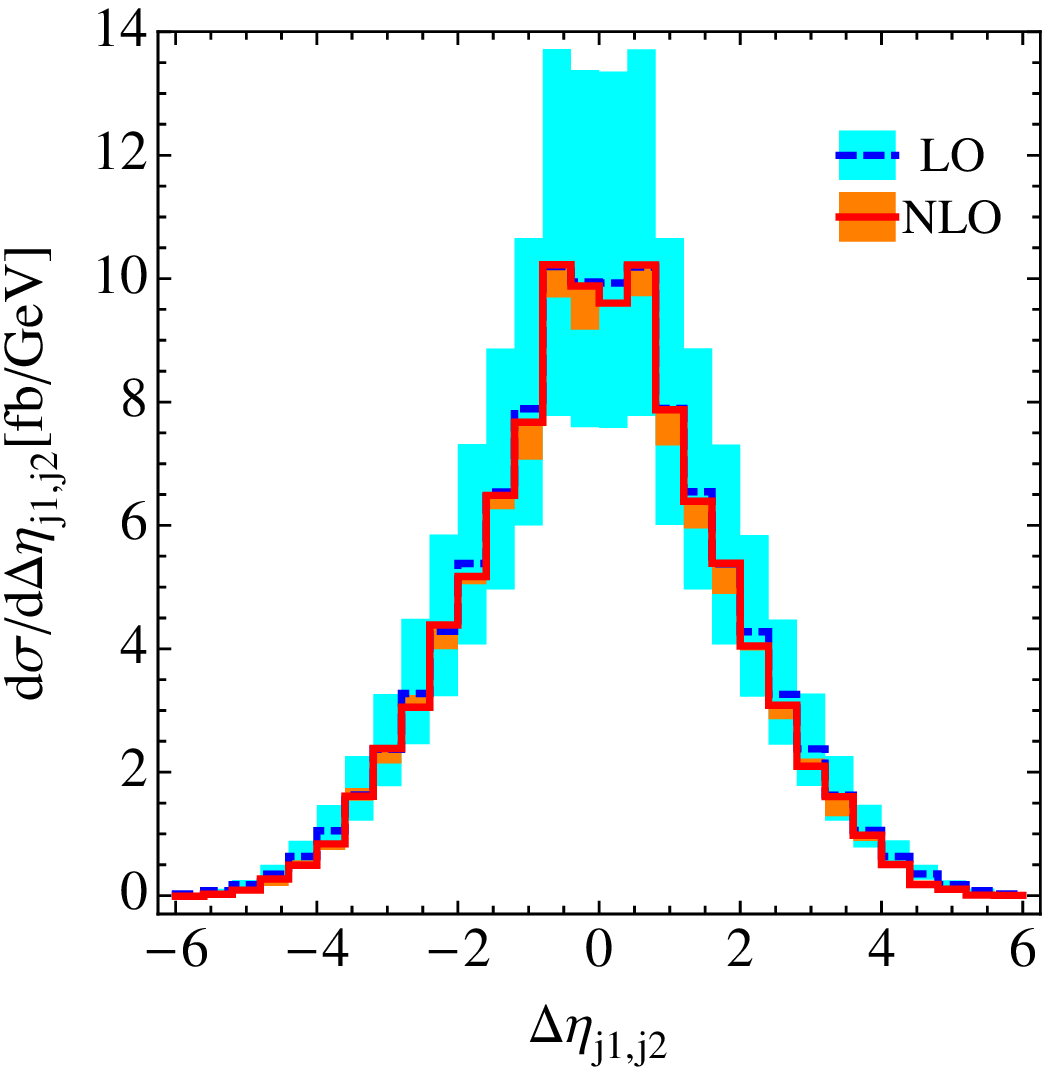}
\qquad
\includegraphics[width=6.5cm]{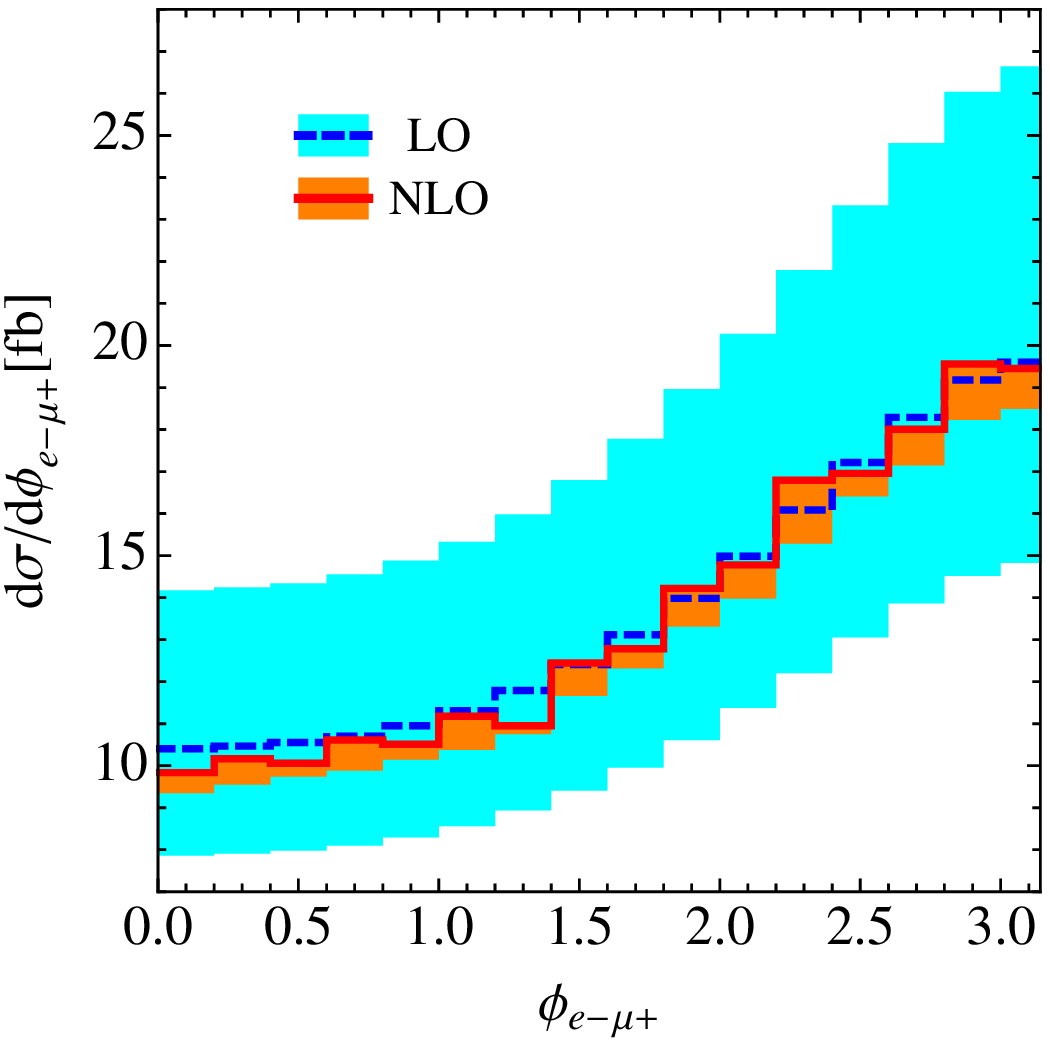}
\vspace{0.2cm}
\includegraphics[width=6.5cm]{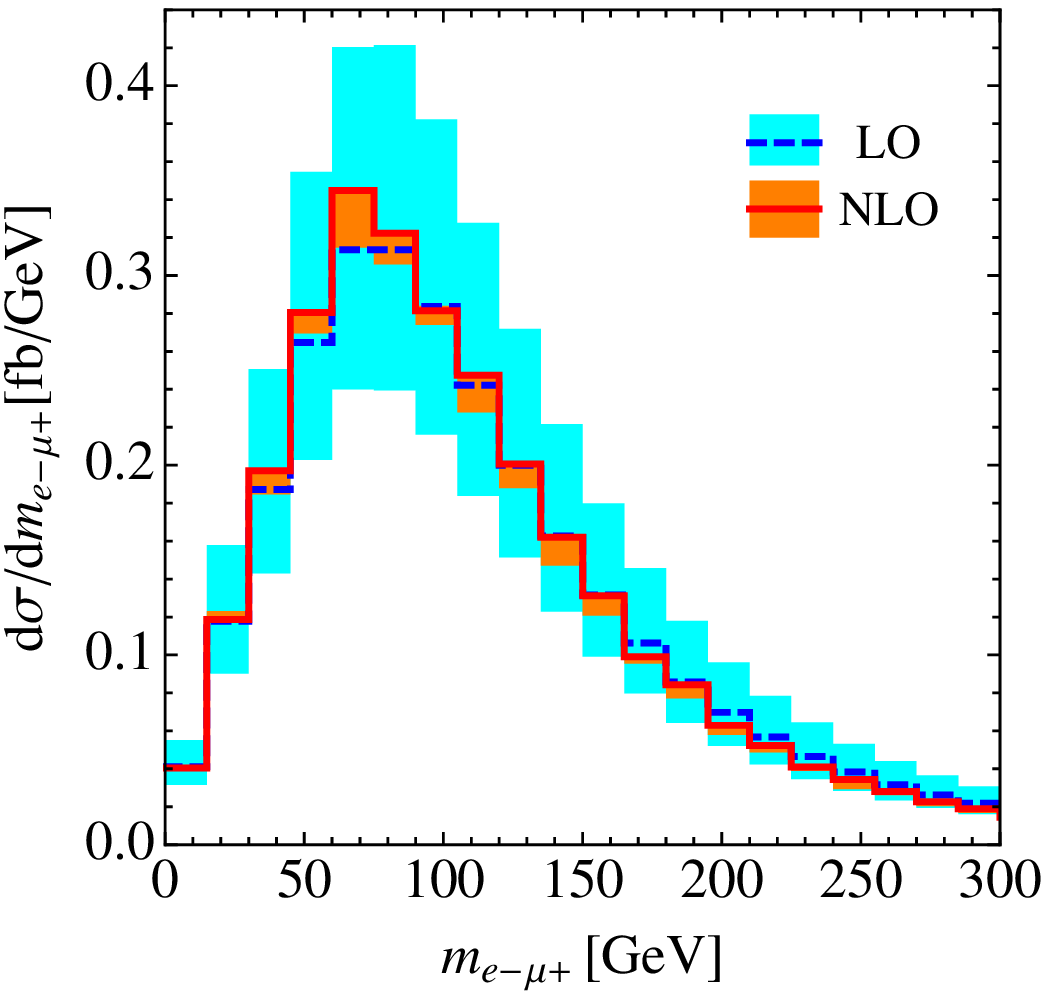}
\qquad
\includegraphics[width=6.5cm]{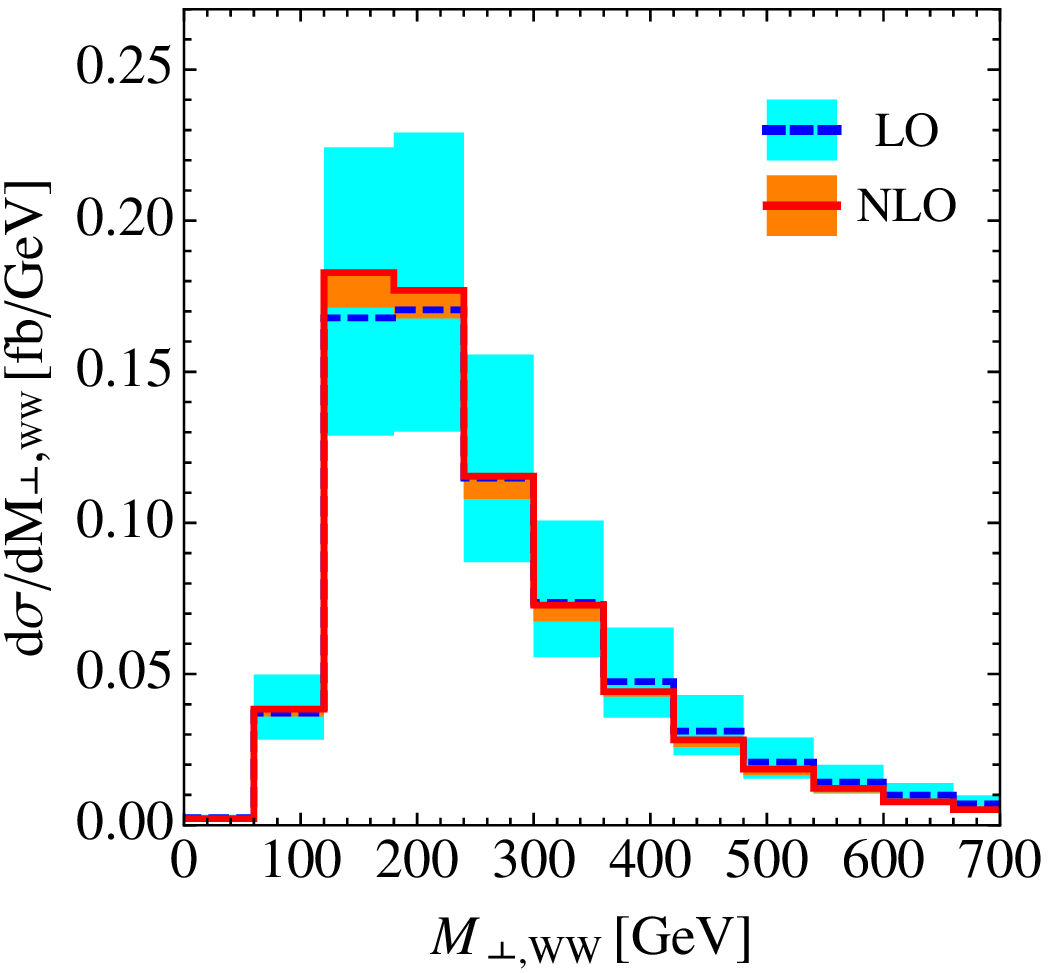}
\caption{Distributions of jet pseudorapidity difference, lepton
  opening angle and invariant masses for $pp \to \nu_{\mu} \mu^+ e^-
  \bar{\nu}_e j j$ at the $7~$TeV run of the LHC.  LO results are
  shown in blue; NLO results are in red and
  orange. The uncertainty bands are for scale $M_W \leq \mu \leq
  4M_W$, and the solid lines show the results at $\mu = 2M_W$.}
\label{figang}
\end{center}
\end{figure}

A few other distributions which are relevant for designing cuts for
Higgs searches are presented in Fig.~\ref{figang}. The distribution of
the relative azimuthal angle between the two leptons is peaked at
$\phi _{e^- \mu^+} = \pi$, with the NLO corrections making almost no
change in the shape of that distribution. The pseudorapidity
difference between two leading jets, defined as $\Delta \eta_{j1,j2} =
\eta_{j1} - \eta_{j2}$, peaks at small values of $\Delta \eta_{j1,j2}$
and falls off rapidly for larger values. The invariant mass of the
leptons and the transverse mass of the $W$-bosons\footnote{We define
  $M_{\perp,WW}^2 = (E_{\perp,l^+l^-} + E_{\perp, \mathrm{miss}})^2 -
  ({\bf p}_{\perp,l^+l^-} - {\bf p}_{\perp,\mathrm{miss}})^2$, where
  $E_{\perp, \mathrm{miss}} = \sqrt{ {\bf p}_{\perp, \mathrm{miss}}^2
    + m_{l^+l^-}^2}$. } become somewhat softer once the NLO QCD
corrections are included.  A discussion of how these distributions can
be used in searches for the Higgs boson can be found in
Refs.~\cite{Campbell:2010cz,Klamke:2007cu,Dittmar:1996ss}.  The
availability of NLO QCD predictions for those distributions should,
potentially, improve the reliability of such analyses since, as
follows from the discussion in this paper, theoretical uncertainties
are reduced considerably.

\section{Conclusions}
In this paper, we computed the NLO QCD corrections to the production
of a $W^+W^-$ pair in association with two jets in hadron collisions.
We only considered the QCD contribution to this process, ignoring the
possibility that it can also occur through exchanges of electroweak
gauge bosons. Our calculation includes the leptonic decays of
$W$-bosons and accounts for all spin correlations exactly.
 
The computation of NLO QCD corrections was performed using the method
of $D$-dimensional generalized unitarity
\cite{Ellis:2007br,Giele:2008ve}.  Practical implementations of the
generalized unitarity technique require color ordering\footnote{See,
  however, a recent discussion in Ref.~\cite{Giele:2009ui}.}; for this
reason, the presence of any colorless particle leads to additional
complication since colorless particles can not be ordered.  Most
processes for which the NLO QCD corrections have been computed using
the on-shell methods involve at most one colorless particle.  The
results of this paper and of Ref.~\cite{Melia:2010bm} show that
generalized unitarity methods can be efficiently used to deal with
processes with a larger number of colorless particles, although the
most general framework for that is yet to be understood.

We studied some phenomenology of the $W^+W^-jj$ production at the Tevatron
and the LHC, using $\sqrt{s} = 7~{\rm TeV}$ for the center-of-mass
collision energy of the latter. We also explored the behaviour of the
NLO QCD cross-section for $pp \to W^+W^-jj$ as a function of the
center-of-mass energy at the LHC and find that, to a good
approximation, the NLO cross-section grows linearly with the energy of
the collider.  For the renormalization and factorization scales set to
$\mu = M_W$ and $\mu = 2M_W$ at the Tevatron and the LHC respectively,
the radiative corrections for both colliders are moderate; in fact
they are very small for collisions at 7 TeV.  We show that the
uncertainty in the theoretical prediction, estimated by changing
factorization and renormalization scales in the range $0.5M_W~ (M_W) <
\mu < 2M_W~ (4 M_W)$ at the Tevatron (LHC) is better than $10\%$ if
the NLO QCD corrections are included.  Of course, at that level of
precision other uncertainties -- such as e.g. the imperfect knowledge
of parton distribution functions -- become important.  We considered a
number of kinematic distributions that involve lepton and jet momenta
and observed that energy-related distributions ($p_\perp$, $H_{\rm
  TOT}$) become softer once the NLO QCD corrections are included and
that shapes of angular distributions are hardly affected.  We also
discussed the significance of $pp \to W^+W^-jj$ process as an
irreducible background for the production of the Higgs boson in
association with two jets at the Tevatron, as well as kinematic
variables useful to disentangle a Higgs signal from the $W^+W^-$
background.

\section*{Acknowledgments}
We are thankful to Keith Ellis and Zoltan Kunszt for their comments on
the manuscript, and to Valentin Hirschi for useful correspondence during
the preparation of this paper.
This research was supported by the NSF under grant PHY-0855365, the
start-up funds provided by the Johns Hopkins University and by the
British Science and Technology Facilities Council, by the LHCPhenoNet
network under the Grant Agreement PITN-GA-2010-264564, and by the
European Research and Training Network (RTN) grant “Unification in the
LHC era ” under the Agreement PITN-GA-2009-237920.  
We would all like to thank CERN for hospitality while part of this
work was carried out.  T.M.~and R.R. would also like to acknowledge
the hospitality extended to them by the Particle Physics Theory Group
at Johns Hopkins University, in the course of the work on this paper.

\newpage

\appendix
\section{Results at a fixed phase space point}
In this Appendix, we shall give numerical results for some of the
tree-level, primitive and full virtual amplitudes used in this
calculation.  For the sake of brevity, amplitudes for some flavors
and helicities are not been reported here. However, we also give
results for squared amplitudes, summed over helicities and color. 

We begin by considering the process $0 \rightarrow (\bar{q} q) +
(W^{+} \rightarrow \nu_{\mu} + \mu^+) + (W^- \rightarrow e^- +
\bar{\nu}_{e}) + g +g$, and use the phase space point defined by the
following values of momenta
\begin{small}
\begin{equation}
 \begin{split}
&p_1^{\bar{u}} = (-500.00000000000000,-500.00000000000000,0.00000000000000,0.00000000000000), \\
&p_2^{u} = (-500.00000000000000,500.00000000000000,0.00000000000000,0.00000000000000), \\
&p_3^{\nu_{\mu}} = (85.5312248384887,-8.22193223977868,36.16378376820329,-77.0725048002413), \\
&p_4^{\mu^+} = (181.42881161004266,-57.85998294819373,-171.863734086635%16
,-5.611858984813%11
), \\
&p_5^{e^-} = (82.84930107743558,-65.90954762358915, -49.89521571962871,5.51413360058664),\\
&p_6^{\bar{\nu}_e} = (381.47038530081545,190.18527704151887,292.042940984587%01
,-155.113300136598%01
), \\
&p_7^{g} = ( 54.23140701179994,-31.13301620817981, -7.9279665679114,43.69128236111634), \\
&p_8^{g} = (214.48887016141776,-27.06079802177751,-98.519808378615,188.59224795994947
). 
 \end{split} \label{appmom}
\end{equation}
\end{small}
Our convention for displaying four-momenta  is $p = (E,p_x,p_y,p_z)$; all momenta are given  
in GeV.

We only include results for the case in which the helicities of the $\bar{u}u$ are $(+,-)$, even though the opposite helicities do contribute via an intermediate vector boson. 
We also do not include the results for $\bar{d}d$ - these can be obtained by switching 
the order of the $W$-bosons and modifying the $\gamma/Z$ couplings in equation 
\eqref{gamZ}. In Tables \ref{tab1}, \ref{tab2}, \ref{tab3} and \ref{tabfermloop}, we give 
tree-level amplitudes as well as the ratios of the unrenormalized virtual amplitudes to the tree-level 
amplitudes
\begin{equation}
\begin{split}
 r_1 = \frac{1}{c_\Gamma}\frac{A_1}{A_0},\;\;\;\; %\\ 
r_1^{[1/2]} = \frac{1}{c_\Gamma}\frac{A_1^{[1/2]}}{A_0}\,,
\end{split}
\end{equation}
where $\displaystyle c_\Gamma = \frac{\Gamma(1+\epsilon)\Gamma^2(1-\epsilon)}{(4\pi)^{2-\epsilon}\Gamma(1-2\epsilon)}$, and the renormalization scale is $\mu_R = 150$ GeV. The tree-level
amplitudes $A_0$ are defined in Eq.~\eqref{ggtree}, 
while the primitive amplitudes $A_1$ are defined in Eq.~\eqref{ggvirt}. The one-loop 
amplitudes are calculated in the four-dimensional helicity scheme \cite{Bern:1991aq,Bern:2002zk}.
Finally, in Table \ref{tab4} we give the ratio 
\begin{equation}
 S_A = \frac{4\pi}{\alpha_s} \frac{\sum_{\mathrm{\{hel\}}} \mathrm{Re}(\mathcal{A}^{\mathrm{tree}} \mathcal{A}^{\mathrm{1L}*})}{\sum_{\mathrm{\{hel\}}}| \mathcal{A}^{\mathrm{tree}} | ^2}\,,
\end{equation}
where the sum is over all helicities for the quarks and gluons.

We now consider the case of $0 \rightarrow (\bar{q} q) + (W^{+} \rightarrow \nu_{\mu} + \mu^+) + (W^- \rightarrow e^- + \bar{\nu_{e}}) + \bar{q}_3 +q_4$. We use the same momenta 
as in equation \eqref{appmom}, with the modification that the last two
momenta in Eq.~(\ref{appmom}) are now those of a $\bar{q}_3 q_4$ pair, 
$p_7^g \to p_7^{\bar{q}_3}$, $p_8^g \to  p_8^{q_4}$. 
For the sake of brevity, we restrict the  results given here to two sets of 
flavors: $\bar{u}u\bar{c}c$ and $\bar{u}d\bar{s}c$ (we are thus working with an ``$s$-amplitude''). The flavor structure of 
the first set is given in Eq.~\eqref{uucc}. We shall also restrict 
ourselves to the helicities $\bar{q}q\bar{q}q = (+,-,+,-)$, although for the former set of flavors, there are four different helicity combinations that are used in the 
calculation. We give the ratios 
\begin{equation}
 r_i = \frac{1}{c_{\Gamma}} \frac{B_1^{(i)}}{B_0}\,,
\end{equation}
for $i=a,b,c,d$, where $B_0$ is defined in Eq.~\eqref{qqtree} and $B_1^{(i)}$ 
are defined in equations \eqref{qqvirt}, \eqref{qqprim1}, and \eqref{qqprim2}. The results
are shown in Tables \ref{tab5} and \ref{tab6}.
We also give the ratios
\begin{equation}
 S_B = \frac{4\pi}{\alpha_s} \frac{ \sum_{\mathrm{\{hel\}}} \mathrm{Re}(\mathcal{B}^{\mathrm{tree}} \mathcal{B}^{\mathrm{1L}*})}{\sum_{\mathrm{\{hel\}}} | \mathcal{B}^{\mathrm{tree}} | ^2}
\end{equation}
in Table \ref{tab7}.

\begin{table}[t]
\begin{center}
\begin{tabular}{|l|c|c|c|}
\hline
Amplitude& $\; 1/\e^2 \;$ & $1/\e$ & $\e^0$ \\
\hline
$\;A_0(\bar{q_1}^+,g_3^-,g_4^-,q_2^-)$&& 
 & $\;  -3.344186 +i\, 9.912207 \;
%\times 10^{-11}\,$GeV$^{-4}\;
$
\\
$\;r_1(\bar{q_1}^+,g_3^-,g_4^-,q_2^-)$
&$\;-1.000000\;$
& $\; 2.294240 -i\,    3.141593$ & $\;    0.4601166 + i\,  2.774496 \;$\\
\hline 
$\;A_0(\bar{q_1}^+,g_3^-,g_4^+,q_2^-)$&& 
 & $\;  0.7055311 +i\,    6.682640 
%\times 10^{-11}\,$GeV$^{-4}\;
$
\\
$\;r_1(\bar{q_1}^+,g_3^-,g_4^+,q_2^-)$
&$\;-1.000000\;$
& $\; 2.294240 -i\,    3.141593$ & $\;    0.3739239 +i\,  2.687541\;$\\
\hline
$\;A_0(\bar{q_1}^+,g_3^+,g_4^-,q_2^-)$&& 
 & $\;  -5.998084 -i\,    5.572010 
%\times 10^{-11}\,$GeV$^{-4}\;
$
\\ 
$\;r_1(\bar{q_1}^+,g_3^+,g_4^-,q_2^-)$
&$\;-1.000000\;$
& $\; 2.294240-i\,    3.141593$ & $\;    0.5484790 +i\,  3.010535\;$\\
\hline 
$\;A_0(\bar{q_1}^+,g_3^+,g_4^+,q_2^-)$&& 
 & $\;  -10.07279 -i\,3.926576    
%\times 10^{-11}\,$GeV$^{-4}\;
$
\\
$\;r_1(\bar{q_1}^+,g_3^+,g_4^+,q_2^-)$
&$\;-1.000000\;$
& $\; 2.294240-i\,    3.141593$ & $\;    0.4741562+i\, 2.846111 \;$\\
\hline 
\end{tabular}
\end{center}
\caption{
  Numerical results for the primitive tree-level amplitude $A_0(\bar{q_1},g_3,g_4,q_2)$, 
  in units of $10^{-10}~{\rm GeV}^{-4}$ 
  and the ratios of primitive one-loop amplitudes $r_1(\bar{q_1},g_3,g_4,q_2)$.
}
\label{tab1}
\end{table}

\begin{table}[t]
\begin{center}
\begin{tabular}{|l|c|c|c|}
\hline
Amplitude& $\; 1/\e^2 \;$ & $1/\e$ & $\e^0$ \\
\hline
$\;A_0(\bar{q_1}^+,g_3^-,q_2^-,g_4^-)$&& 
 & $\;  -5.097350 +i\,    3.386328 
%\times 10^{-11}\,$GeV$^{-4}\;
$
\\
$\;r_1(\bar{q_1}^+,g_3^-,q_2^-,g_4^-)$
&$\;-2.000000\;$
& $\; 2.993440+i\,    0.000000$ & $\;    -0.07739397+i\,  3.420824\;$\\
\hline 
$\;A_0(\bar{q_1}^+,g_3^-,q_2^-,g_4^+)$&& 
 & $\;  -4.426865 +i\,    4.803504 
%\times 10^{-11}\,$GeV$^{-4}\;
$
\\
$\;r_1(\bar{q_1}^+,g_3^-,q_2^-,g_4^+)$
&$\;-2.000000\;$
& $\; 2.993440+i\,    0.000000$ & $\;    6.347479+i\,  5.196425\;$\\
\hline
$\;A_0(\bar{q_1}^+,g_3^+,q_2^-,g_4^-)$&& 
 & $\;  -4.749089 +i\, 1.306764   
%\times 10^{-11}\,$GeV$^{-4}\;
$
\\ 
$\;r_1(\bar{q_1}^+,g_3^+,q_2^-,g_4^-)$
&$\;-2.000000\;$
& $\; 2.993440+i\,    0.000000$ & $\;    -0.8538774+i\,  3.373345\;$\\
\hline 
$\;A_0(\bar{q_1}^+,g_3^+,q_2^-,g_4^+)$&& 
 & $\;  -8.206743 +i\, 2.583236     
%\times 10^{-11}\,$GeV$^{-4}\;
$
\\
$\;r_1(\bar{q_1}^+,g_3^+,q_2^-,g_4^+)$
&$\;-2.000000\;$
& $\; 2.993440+i\,    0.000000$ & $\;    6.051784+i\,  4.612948\;$\\
\hline 
\end{tabular}
\end{center}
\caption{
  Numerical results for the primitive tree-level amplitude $A_0(\bar{q_1},g_3,q_2,g_4)$, 
  in units of $10^{-10}~{\rm GeV}^{-4}$ 
  and the ratios of primitive one-loop amplitudes $r_1(\bar{q_1},g_3,q_2,g_4)$.
}
\label{tab2}
\end{table}

\newpage

\begin{table}[t]
\begin{center}
\begin{tabular}{|l|c|c|c|}
\hline
Amplitude& $\; 1/\e^2 \;$ & $1/\e$ & $\e^0$ \\
\hline
$\;A_0(\bar{q_1}^+,q_2^-,g_3^-,g_4^-)$&& 
 & $\;  8.441536 -i\,    13.29854 
%\times 10^{-11}\,$GeV$^{-4}\;
$
\\
$\;r_1(\bar{q_1}^+,q_2^-,g_3^-,g_4^-)$
&$\;-3.000000\;$
& $\; -0.9503441-i\,    3.141593$ & $\;    -6.047837-i\, 9.654414\;$\\
\hline 
$\;A_0(\bar{q_1}^+,q_2^-,g_3^-,g_4^+,)$&& 
 & $\;  3.721334 -i\, 11.48614    
%\times 10^{-11}\,$GeV$^{-4}\;
$
\\
$\;r_1(\bar{q_1}^+,q_2^-,g_3^-,g_4^+)$
&$\;-3.000000\;$
& $\; -0.9503441-i\,    3.141593$ & $\;    0.9335325 -i\,  8.464906\;$\\
\hline
$\;A_0(\bar{q_1}^+,q_2^-,g_3^+,g_4^-)$&& 
 & $\;  10.74717 +i\ 4.265245,     
%\times 10^{-11}\,$GeV$^{-4}\;
$
\\ 
$\;r_1(\bar{q_1}^+,q_2^-,g_3^+,g_4^-)$
&$\;-3.000000\;$
& $\; -0.9503441 -i\,    3.141593$ & $\;    -6.036407-i\,  10.58605\;$\\
\hline 
$\;A_0(\bar{q_1}^+,q_2^-,g_3^+,g_4^+)$&& 
 & $\;  18.27953 +i\, 1.343340     
%\times 10^{-11}\,$GeV$^{-4}\;
$
\\
$\;r_1(\bar{q_1}^+,q_2^-g_3^+,g_4^+)$
&$\;-3.000000\;$
& $\; -0.9503441-i\,    3.141593$ & $\;    0.3979266-i\, 9.181091\;$\\
\hline 
\end{tabular}
\end{center}
\caption{
  Numerical results for the primitive tree-level amplitude $A_0(\bar{q_1},q_2,g_3,g_4)$, 
  in units of $10^{-10}~{\rm GeV}^{-4}$ 
  and the ratios of primitive one-loop amplitudes $r_1(\bar{q_1},q_2,g_3,g_4)$.
}
\label{tab3}
\end{table}

\vspace*{1.5cm}

\begin{table}[h]
\begin{center}
\begin{tabular}{|l|c|}
\hline
Amplitude& $\; \e^0$ \\
\hline
$\;A_0(\bar{q_1}^+,q_2^-,g_3^-,g_4^-)$&
  $\;  8.441536 -i\,    13.29854 
%\times 10^{-11}\,$GeV$^{-4}\;
$
\\
$\;r_1^{[1/2]}(\bar{q_1}^+,q_2^-,g_3^-,g_4^-)$&
$\;    (-0.3523178 -i\, 4.071390) \times 10^{-2} \;$\\
\hline 
$\;A_0(\bar{q_1}^+,q_2^-,g_3^-,g_4^+,)$&
  $\;  3.721334 -i\, 11.48614    
%\times 10^{-11}\,$GeV$^{-4}\;
$
\\
$\;r_1^{[1/2]}(\bar{q_1}^+,q_2^-,g_3^-,g_4^+)$
&$\; 0.000000 +i\ 0.000000\;$\\
\hline
$\;A_0(\bar{q_1}^+,q_2^-,g_3^+,g_4^-)$ &
  $ \; 10.74717 +i\ 4.265245    
%\times 10^{-11}\,$GeV$^{-4}\;
$
\\ 
$\;r_1^{[1/2]}(\bar{q_1}^+,q_2^-,g_3^+,g_4^-)$
&$\;    0.000000 +i\,  0.000000\;$\\
\hline 
$\;A_0(\bar{q_1}^+,q_2^-,g_3^+,g_4^+)$
 & $\;  18.27953 +i\, 1.343340     
%\times 10^{-11}\,$GeV$^{-4}\;
$
\\
$\;r_1^{[1/2]}(\bar{q_1}^+,q_2^-g_3^+,g_4^+)$
&$\; (-3.142652 +i\,1.567695) \times 10^{-2} \;$\\
\hline 
\end{tabular}
\end{center}
\caption{
  Numerical results for the primitive tree-level amplitude $A_0(\bar{q_1},q_2,g_3,g_4)$, 
  in units of $10^{-10}~{\rm GeV}^{-4}$ 
  and the ratios of primitive one-loop amplitudes $r_1^{[1/2]}(\bar{q_1},q_2,g_3,g_4)$. There are no singular contributions from these one-loop amplitudes, and swapping the gluons simply changes the sign of the amplitude.
}
\label{tabfermloop}
\end{table}

\vspace*{1.5cm}

\begin{table}[h]
\begin{center}
\begin{tabular}{|l|c|c|c|}
\hline
Ratio& $\; 1/\e^2 \;$ & $1/\e$ & $\e^0$ \\
\hline
$\;\sum| \mathcal{A}^{\mathrm{tree}}(\bar{d},d,g,g) | ^2$
&$\;\;$
& $\; $ & $\; 9.887737 \times 10^{-20}\;$\\
\hline
$\;S_A(\bar{d},d,g,g)$
&$\;-8.666667\;$
& $\; -2.836720$ & $\;  -0.6913131\;$\\
\hline 
$\;\sum| \mathcal{A}^{\mathrm{tree}}(\bar{u},u,g,g) | ^2$
&$\;\;$
& $\; $ & $\; 3.743231 \times 10^{-20}\;$\\
\hline
$\;S_A(\bar{u},u,g,g)$
&$\;-8.666667\;$
& $\; -2.786885$ & $\;   -4.673601 \;$\\

\hline 
\end{tabular}
\end{center}
\caption{Numerical results for the tree-level amplitude squared, in units of GeV$^{-8}$, and the ratio of virtual over tree-level
  squared amplitudes $S_A$ summed over all helicities and colors.}

\label{tab4}
\end{table}

\newpage

\begin{table}[h]
\begin{center}
\begin{tabular}{|l|c|c|c|}
\hline
Amplitude& $\; 1/\e^2 \;$ & $1/\e$ & $\e^0$ \\
\hline
$\;B_0(\bar{u},u,\bar{c},c)$&& 
 & $\;  0.6391654 +i\, 5.544406 
%\times 10^{-11}\,$GeV$^{-4}\;
$
\\
$\;r_a(\bar{u},u,\bar{c},c)$
&$\;-2.000000\;$
& $\; 3.066474 + i\, 0.000000$ & $\;   2.658086+i\, 2.684586\;$\\
$\;r_a(\bar{u},u,c,\bar{c})$
&$\;-2.000000\;$
& $\; 4.119961 +i\, 0.000000$ & $\;    3.634715+i\, 2.090514\;$\\
$\;r_b(\bar{u},u,\bar{c},c)$
&$\;-1.000000\;$
& $\; 2.294240-i\,    3.141593$ & $\;    0.1918562 + i\, 2.854994\;$\\
$\;r_c(\bar{u},u,\bar{c},c)$
&$\;-1.000000\;$
& $\; -3.350152-i\,    3.141593$ & $\;    -3.028899 -i\,10.77523\;$\\
$\;r_d(\bar{u},u,\bar{c},c)$
&$\;\;$
& $\; -0.6666667 + i\, 0.000000$ & $\; -2.301323 - i\,1.838568   \;$\\
\hline 
\end{tabular}
\end{center}
\caption{
  Numerical results for the primitive tree-level amplitude $B_0(\bar{u},u,\bar{c},c)$, 
  in units of $10^{-10}~{\rm GeV}^{-4}$ 
  and the ratios of primitive one-loop amplitudes $r_i$.
}
\label{tab5}
\end{table}

\vspace*{1.5cm}

\begin{table}[h]
\begin{center}
\begin{tabular}{|l|c|c|c|}
\hline
Amplitude& $\; 1/\e^2 \;$ & $1/\e$ & $\e^0$ \\
\hline
$\;B_0(\bar{u},d,\bar{s},c)$&& 
 & $\;  0.3350897 -i\, 0.6484033 
%\times 10^{-11}\,$GeV$^{-4}\;
$
\\
$\;r_a(\bar{u},d,\bar{s},c)$
&$\;-2.000000\;$
& $\; 3.066474 + i\, 0.000000$ & $\;   -7.426922 -i\, 0.3913681\;$\\
$\;r_a(\bar{u},d,c,\bar{s})$
&$\;-2.000000\;$
& $\; 4.119961 +i\, 0.000000$ & $\;    -7.135027-i\, 13.92234\;$\\
$\;r_b(\bar{u},d,\bar{s},c)$
&$\;-1.000000\;$
& $\; 2.294240-i\,    3.141593$ & $\;    0.7221019 + i\, 6.182924\;$\\
$\;r_c(\bar{u},d,\bar{s},c)$
&$\;-1.000000\;$
& $\; -3.350152-i\,    3.141593$ & $\;    -7.635220 -i\,10.639296\;$\\
$\;r_d(\bar{u},d,\bar{s},c)$
&$\;\;$
& $\; -0.6666667 + i\, 0.000000$ & $\; 0.9665113 - i\, 2.094395   \;$\\
\hline 
\end{tabular}
\end{center}
\caption{
  Numerical results for the primitive tree-level amplitude $B_0(\bar{u},d,\bar{s},c)$, 
  in units of $10^{-10}~{\rm GeV}^{-4}$ 
  and the ratios of primitive one-loop amplitudes $r_i$.
}
\label{tab6}
 \end{table}

\vspace*{1.5cm}

\begin{table}[h]
\begin{center}
\begin{tabular}{|l|c|c|c|}
\hline
Ratio& $\; 1/\e^2 \;$ & $1/\e$ & $\e^0$ \\
\hline
$\;\sum | \mathcal{B}^{\mathrm{tree}} (\bar{u},u,\bar{c},c)| ^2$
&$\;\;$
& $\;$ & $\;  1.037139 \times 10^{-21}\;$\\
\hline
$\;S_B(\bar{u},u,\bar{c},c)$
&$\;-5.333333\;$
& $\; 7.587051$ & $\;  5.395242\;$\\
\hline 
$\;\sum | \mathcal{B}^{\mathrm{tree}} (\bar{u},d,\bar{s},c)| ^2$
&$\;\;$
& $\;$ & $\;  1.123763\times 10^{-23}\;$\\
\hline
$\;S_B(\bar{u},d,\bar{s},c)$
&$\;-5.333333\;$
& $\; 7.587051$ & $\;   -15.91575 \;$\\

\hline 
\end{tabular}
\end{center}
\caption{Numerical results for tree-level amplitudes squared, in units of GeV$^{-8}$, and the ratio of virtual over tree-level
  squared amplitudes $S_B$ summed over all helicities and colors.}
\label{tab7}
\end{table}


\newpage 

\begin{thebibliography}{100}

\bibitem{Denner:2002ii}
A.~Denner and S.~Dittmaier, {\it {Reduction of one-loop tensor 5-point
  integrals}},  {\em Nucl. Phys.} {\bf B658} (2003) 175--202,
  [\href{http://xxx.lanl.gov/abs/hep-ph/0212259}{{\tt hep-ph/0212259}}].

\bibitem{Binoth:2005ff}
T.~Binoth, J.~P. Guillet, G.~Heinrich, E.~Pilon, and C.~Schubert, {\it {An
  algebraic / numerical formalism for one-loop multi-leg amplitudes}},  {\em
  JHEP} {\bf 10} (2005) 015,
  [\href{http://xxx.lanl.gov/abs/hep-ph/0504267}{{\tt hep-ph/0504267}}].

\bibitem{Denner:2005nn}
A.~Denner and S.~Dittmaier, {\it {Reduction schemes for one-loop tensor
  integrals}},  {\em Nucl. Phys.} {\bf B734} (2006) 62--115,
  [\href{http://xxx.lanl.gov/abs/hep-ph/0509141}{{\tt hep-ph/0509141}}].

\bibitem{Britto:2004nc}
R.~Britto, F.~Cachazo, and B.~Feng, {\it {Generalized unitarity and one-loop
  amplitudes in N = 4 super-Yang-Mills}},  {\em Nucl. Phys.} {\bf B725} (2005)
  275--305, [\href{http://xxx.lanl.gov/abs/hep-th/0412103}{{\tt
  hep-th/0412103}}].

\bibitem{Britto:2004tx}
R.~Britto, F.~Cachazo, and B.~Feng, {\it {Coplanarity in twistor space of N = 4
  next-to-MHV one-loop amplitude coefficients}},  {\em Phys. Lett.} {\bf B611}
  (2005) 167--172, [\href{http://xxx.lanl.gov/abs/hep-th/0411107}{{\tt
  hep-th/0411107}}].

\bibitem{Ossola:2006us}
G.~Ossola, C.~G. Papadopoulos, and R.~Pittau, {\it {Reducing full one-loop
  amplitudes to scalar integrals at the integrand level}},  {\em Nucl. Phys.}
  {\bf B763} (2007) 147--169,
  [\href{http://xxx.lanl.gov/abs/hep-ph/0609007}{{\tt hep-ph/0609007}}].

\bibitem{Forde:2007mi}
D.~Forde, {\it {Direct extraction of one-loop integral coefficients}},  {\em
  Phys. Rev.} {\bf D75} (2007) 125019,
  [\href{http://xxx.lanl.gov/abs/0704.1835}{{\tt arXiv:0704.1835}}].

\bibitem{Ellis:2007br}
R.~K. Ellis, W.~T. Giele, and Z.~Kunszt, {\it {A Numerical Unitarity Formalism
  for Evaluating One-Loop Amplitudes}},  {\em JHEP} {\bf 03} (2008) 003,
  [\href{http://xxx.lanl.gov/abs/0708.2398}{{\tt arXiv:0708.2398}}].

\bibitem{Giele:2008ve}
W.~T. Giele, Z.~Kunszt, and K.~Melnikov, {\it {Full one-loop amplitudes from
  tree amplitudes}},  {\em JHEP} {\bf 04} (2008) 049,
  [\href{http://xxx.lanl.gov/abs/0801.2237}{{\tt arXiv:0801.2237}}].

\bibitem{Ossola:2008xq}
G.~Ossola, C.~G. Papadopoulos, and R.~Pittau, {\it {On the Rational Terms of
  the one-loop amplitudes}},  {\em JHEP} {\bf 05} (2008) 004,
  [\href{http://xxx.lanl.gov/abs/0802.1876}{{\tt arXiv:0802.1876}}].

\bibitem{Berger:2009ep}
C.~F. Berger {\em et.~al.}, {\it {Next-to-Leading Order QCD Predictions for
  W+3-Jet Distributions at Hadron Colliders}},  {\em Phys. Rev.} {\bf D80}
  (2009) 074036, [\href{http://xxx.lanl.gov/abs/0907.1984}{{\tt
  arXiv:0907.1984}}].

\bibitem{Berger:2009zg}
C.~F. Berger {\em et.~al.}, {\it {Precise Predictions for $W$ + 3 Jet
  Production at Hadron Colliders}},  {\em Phys. Rev. Lett.} {\bf 102} (2009)
  222001, [\href{http://xxx.lanl.gov/abs/0902.2760}{{\tt arXiv:0902.2760}}].

\bibitem{KeithEllis:2009bu}
R.~K. Ellis, K.~Melnikov, and G.~Zanderighi, {\it {W+3 jet production at the
  Tevatron}},  {\em Phys. Rev.} {\bf D80} (2009) 094002,
  [\href{http://xxx.lanl.gov/abs/0906.1445}{{\tt arXiv:0906.1445}}].

\bibitem{Ellis:2009zw}
R.~K. Ellis, K.~Melnikov, and G.~Zanderighi, {\it {Generalized unitarity at
  work: first NLO QCD results for hadronic $W^+$ 3jet production}},  {\em JHEP}
  {\bf 04} (2009) 077, [\href{http://xxx.lanl.gov/abs/0901.4101}{{\tt
  arXiv:0901.4101}}].

\bibitem{Melnikov:2009wh}
K.~Melnikov and G.~Zanderighi, {\it {W+3 jet production at the LHC as a signal
  or background}},  {\em Phys. Rev.} {\bf D81} (2010) 074025,
  [\href{http://xxx.lanl.gov/abs/0910.3671}{{\tt arXiv:0910.3671}}].

\bibitem{Berger:2010vm}
C.~F. Berger {\em et.~al.}, {\it {Next-to-Leading Order QCD Predictions for
  Z,$\gamma^*$+3-Jet Distributions at the Tevatron}},  {\em Phys. Rev.} {\bf
  D82} (2010) 074002, [\href{http://xxx.lanl.gov/abs/1004.1659}{{\tt
  arXiv:1004.1659}}].

\bibitem{Bredenstein:2009aj}
A.~Bredenstein, A.~Denner, S.~Dittmaier, and S.~Pozzorini, {\it {NLO QCD
  corrections to pp $\to$ t anti-t b anti-b + X at the LHC}},  {\em Phys. Rev.
  Lett.} {\bf 103} (2009) 012002,
  [\href{http://xxx.lanl.gov/abs/0905.0110}{{\tt arXiv:0905.0110}}].

\bibitem{Bredenstein:2010rs}
A.~Bredenstein, A.~Denner, S.~Dittmaier, and S.~Pozzorini, {\it {NLO QCD
  corrections to top anti-top bottom anti-bottom production at the LHC: 2. full
  hadronic results}},  {\em JHEP} {\bf 03} (2010) 021,
  [\href{http://xxx.lanl.gov/abs/1001.4006}{{\tt arXiv:1001.4006}}].

\bibitem{Bevilacqua:2009zn}
G.~Bevilacqua, M.~Czakon, C.~G. Papadopoulos, R.~Pittau, and M.~Worek, {\it
  {Assault on the NLO Wishlist: pp $\rightarrow$ tt bb}},  {\em JHEP} {\bf 09}
  (2009) 109, [\href{http://xxx.lanl.gov/abs/0907.4723}{{\tt
  arXiv:0907.4723}}].

\bibitem{Bevilacqua:2010ve}
G.~Bevilacqua, M.~Czakon, C.~G. Papadopoulos, and M.~Worek, {\it {Dominant QCD
  Backgrounds in Higgs Boson Analyses at the LHC: A Study of pp $\rightarrow$ t
  anti-t + 2 jets at Next-To-Leading Order}},  {\em Phys. Rev. Lett.} {\bf 104}
  (2010) 162002, [\href{http://xxx.lanl.gov/abs/1002.4009}{{\tt
  arXiv:1002.4009}}].

\bibitem{Binoth:2009rv}
T.~Binoth {\em et.~al.}, {\it {Next-to-leading order QCD corrections to $pp -->
  b \bar{b} b \bar{b} + X$ at the LHC: the quark induced case}},  {\em Phys.
  Lett.} {\bf B685} (2010) 293--296,
  [\href{http://xxx.lanl.gov/abs/0910.4379}{{\tt arXiv:0910.4379}}].

\bibitem{Bevilacqua:2010qb}
G.~Bevilacqua, M.~Czakon, A.~van Hameren, C.~G. Papadopoulos, and M.~Worek,
  {\it {Complete off-shell effects in top quark pair hadroproduction with
  leptonic decay at next-to-leading order}},
  \href{http://xxx.lanl.gov/abs/1012.4230}{{\tt arXiv:1012.4230}}.

\bibitem{Denner:2010jp}
A.~Denner, S.~Dittmaier, S.~Kallweit, and S.~Pozzorini, {\it {NLO QCD
  corrections to WWbb production at hadron colliders}},
  \href{http://xxx.lanl.gov/abs/1012.3975}{{\tt arXiv:1012.3975}}.

\bibitem{Melia:2010bm}
T.~Melia, K.~Melnikov, R.~Rontsch, and G.~Zanderighi, {\it {Next-to-leading
  order QCD predictions for W+W+jj production at the LHC}},  {\em JHEP} {\bf
  12} (2010) 053, [\href{http://xxx.lanl.gov/abs/1007.5313}{{\tt
  arXiv:1007.5313}}].

\bibitem{Melia:2011gk}
  T.~Melia, P.~Nason, R.~Rontsch, G.~Zanderighi,
  {\it $W^+W^+$ plus dijet production in the POWHEGBOX},
  [\href{http://xxx.lanl.gov/abs/1102.4846}{{\tt arXiv:1102.4846}}].

\bibitem{Berger:2010zx}
C.~F. Berger {\em et.~al.}, {\it {Precise Predictions for W + 4 Jet Production
  at the Large Hadron Collider}},
  \href{http://xxx.lanl.gov/abs/1009.2338}{{\tt arXiv:1009.2338}}.

\bibitem{Mastrolia:2010nb}
P.~Mastrolia, G.~Ossola, T.~Reiter, and F.~Tramontano, {\it {Scattering
  AMplitudes from Unitarity-based Reduction Algorithm at the Integrand-level}},
   {\em JHEP} {\bf 08} (2010) 080,
  [\href{http://xxx.lanl.gov/abs/1006.0710}{{\tt arXiv:1006.0710}}].

\bibitem{Hirschi:2011pa}
V.~Hirschi, R.~Frederix, S.~Frixione, M.~V. Garzelli, F.~Maltoni, {\em
  et.~al.}, {\it {Automation of one-loop QCD corrections}},
  \href{http://xxx.lanl.gov/abs/1103.0621}{{\tt arXiv:1103.0621}}.

\bibitem{higgscdf}
{\bf CDF} Collaboration, {\it {Search for $H \to WW*$ production at CDF using
  4.8 fb$^{-1}$ of data}}, . CDF note 9887.

\bibitem{Campbell:2006xx}
J.~M. Campbell, R.~K. Ellis, and G.~Zanderighi, {\it {Next-to-leading order
  Higgs + 2 jet production via gluon fusion}},  {\em JHEP} {\bf 10} (2006) 028,
  [\href{http://xxx.lanl.gov/abs/hep-ph/0608194}{{\tt hep-ph/0608194}}].

\bibitem{Anastasiou:2009bt}
C.~Anastasiou, G.~Dissertori, M.~Grazzini, F.~Stockli, and B.~R. Webber, {\it
  {Perturbative QCD effects and the search for a $H \to WW \to l nu l nu$
  signal at the Tevatron}},  {\em JHEP} {\bf 08} (2009) 099,
  [\href{http://xxx.lanl.gov/abs/0905.3529}{{\tt arXiv:0905.3529}}].

\bibitem{Berger:2004pca}
E.~L. Berger and J.~M. Campbell, {\it {Higgs boson production in weak boson
  fusion at next-to- leading order}},  {\em Phys. Rev.} {\bf D70} (2004)
  073011, [\href{http://xxx.lanl.gov/abs/hep-ph/0403194}{{\tt
  hep-ph/0403194}}].

\bibitem{Han:1992hr}
T.~Han, G.~Valencia, and S.~Willenbrock, {\it {Structure function approach to
  vector boson scattering in p p collisions}},  {\em Phys. Rev. Lett.} {\bf 69}
  (1992) 3274--3277, [\href{http://xxx.lanl.gov/abs/hep-ph/9206246}{{\tt
  hep-ph/9206246}}].

\bibitem{Figy:2003nv}
T.~Figy, C.~Oleari, and D.~Zeppenfeld, {\it {Next-to-leading order jet
  distributions for Higgs boson production via weak-boson fusion}},  {\em Phys.
  Rev.} {\bf D68} (2003) 073005,
  [\href{http://xxx.lanl.gov/abs/hep-ph/0306109}{{\tt hep-ph/0306109}}].

\bibitem{Dixon:1999di}
L.~J. Dixon, Z.~Kunszt, and A.~Signer, {\it {Vector boson pair production in
  hadronic collisions at order $\alpha_s$ : Lepton correlations and anomalous
  couplings}},  {\em Phys. Rev.} {\bf D60} (1999) 114037,
  [\href{http://xxx.lanl.gov/abs/hep-ph/9907305}{{\tt hep-ph/9907305}}].

\bibitem{Campbell:1999ah}
J.~M. Campbell and R.~K. Ellis, {\it {An update on vector boson pair production
  at hadron colliders}},  {\em Phys. Rev.} {\bf D60} (1999) 113006,
  [\href{http://xxx.lanl.gov/abs/hep-ph/9905386}{{\tt hep-ph/9905386}}].

\bibitem{Ohnemus:1991gb}
J.~Ohnemus, {\it {An Order $\alpha_s$ calculation of hadronic $W^\pm Z$
  production}},  {\em Phys. Rev.} {\bf D44} (1991) 3477--3489.

\bibitem{Campbell:2007ev}
J.~M. Campbell, K.~R. Ellis, and G.~Zanderighi, {\it {Next-to-leading order
  predictions for $WW+1$ jet distributions at the LHC}},  {\em JHEP} {\bf 12}
  (2007) 056, [\href{http://xxx.lanl.gov/abs/0710.1832}{{\tt
  arXiv:0710.1832}}].

\bibitem{Dittmaier:2009un}
S.~Dittmaier, S.~Kallweit, and P.~Uwer, {\it {NLO QCD corrections to pp/ppbar
  $\to$ WW+jet+X including leptonic W-boson decays}},  {\em Nucl.Phys.} {\bf
  B826} (2010) 18--70, [\href{http://xxx.lanl.gov/abs/0908.4124}{{\tt
  arXiv:0908.4124}}].

\bibitem{Mangano:1990by}
M.~L. Mangano and S.~J. Parke, {\it {Multiparton amplitudes in gauge
  theories}},  {\em Phys.Rept.} {\bf 200} (1991) 301--367,
  [\href{http://xxx.lanl.gov/abs/hep-th/0509223}{{\tt hep-th/0509223}}].

\bibitem{Bern:1994fz}
Z.~Bern, L.~J. Dixon, and D.~A. Kosower, {\it {One loop corrections to two
  quark three gluon amplitudes}},  {\em Nucl. Phys.} {\bf B437} (1995)
  259--304, [\href{http://xxx.lanl.gov/abs/hep-ph/9409393}{{\tt
  hep-ph/9409393}}].

\bibitem{Bern:1996je}
Z.~Bern, L.~J. Dixon, and D.~A. Kosower, {\it {Progress in one-loop QCD
  computations}},  {\em Ann. Rev. Nucl. Part. Sci.} {\bf 46} (1996) 109--148,
  [\href{http://xxx.lanl.gov/abs/hep-ph/9602280}{{\tt hep-ph/9602280}}].

\bibitem{Bern:1997sc}
Z.~Bern, L.~J. Dixon, and D.~A. Kosower, {\it {One-loop amplitudes for e+ e- to
  four partons}},  {\em Nucl. Phys.} {\bf B513} (1998) 3--86,
  [\href{http://xxx.lanl.gov/abs/hep-ph/9708239}{{\tt hep-ph/9708239}}].

\bibitem{Ellis:2008qc}
R.~K. Ellis, W.~T. Giele, Z.~Kunszt, K.~Melnikov, and G.~Zanderighi, {\it
  {One-loop amplitudes for $W^+$ + 3 jet production in hadron collisions}},
  {\em JHEP} {\bf 01} (2009) 012,
  [\href{http://xxx.lanl.gov/abs/0810.2762}{{\tt arXiv:0810.2762}}].

\bibitem{Berends:1987me}
F.~A. Berends and W.~Giele, {\it {Recursive Calculations for Processes with n
  Gluons}},  {\em Nucl.Phys.} {\bf B306} (1988) 759.

\bibitem{Catani:1996vz}
S.~Catani and M.~H. Seymour, {\it {A general algorithm for calculating jet
  cross sections in NLO QCD}},  {\em Nucl. Phys.} {\bf B485} (1997) 291--419,
  [\href{http://xxx.lanl.gov/abs/hep-ph/9605323}{{\tt hep-ph/9605323}}].

\bibitem{Nagy:1998bb}
Z.~Nagy and Z.~Trocsanyi, {\it {Next-to-leading order calculation of four-jet
  observables in electron positron annihilation}},  {\em Phys. Rev.} {\bf D59}
  (1999) 014020, [\href{http://xxx.lanl.gov/abs/hep-ph/9806317}{{\tt
  hep-ph/9806317}}].

\bibitem{Nagy:2003tz}
Z.~Nagy, {\it {Next-to-leading order calculation of three jet observables in
  hadron hadron collision}},  {\em Phys. Rev.} {\bf D68} (2003) 094002,
  [\href{http://xxx.lanl.gov/abs/hep-ph/0307268}{{\tt hep-ph/0307268}}].

\bibitem{Campbell:2000bg}
J.~M. Campbell and R.~K. Ellis, {\it {Radiative corrections to Z b anti-b
  production}},  {\em Phys. Rev.} {\bf D62} (2000) 114012,
  [\href{http://xxx.lanl.gov/abs/hep-ph/0006304}{{\tt hep-ph/0006304}}].

\bibitem{Ellis:2007qk}
R.~Ellis and G.~Zanderighi, {\it {Scalar one-loop integrals for QCD}},  {\em
  JHEP} {\bf 0802} (2008) 002, [\href{http://xxx.lanl.gov/abs/0712.1851}{{\tt
  arXiv:0712.1851}}].

\bibitem{Ellis:2008ir}
R.~Ellis, W.~T. Giele, Z.~Kunszt, and K.~Melnikov, {\it {Masses, fermions and
  generalized D-dimensional unitarity}},  {\em Nucl.Phys.} {\bf B822} (2009)
  270--282, [\href{http://xxx.lanl.gov/abs/0806.3467}{{\tt arXiv:0806.3467}}].

\bibitem{Melnikov:2010iu}
K.~Melnikov and M.~Schulze, {\it {NLO QCD corrections to top quark pair
  production in association with one hard jet at hadron colliders}},  {\em
  Nucl.Phys.} {\bf B840} (2010) 129--159,
  [\href{http://xxx.lanl.gov/abs/1004.3284}{{\tt arXiv:1004.3284}}].

\bibitem{Binoth:2006mf}
T.~Binoth, M.~Ciccolini, N.~Kauer, and M.~Kramer, {\it {Gluon-induced W-boson
  pair production at the LHC}},  {\em JHEP} {\bf 0612} (2006) 046,
  [\href{http://xxx.lanl.gov/abs/hep-ph/0611170}{{\tt hep-ph/0611170}}].

\bibitem{Jager:2006zc}
B.~Jager, C.~Oleari, and D.~Zeppenfeld, {\it {Next-to-leading order QCD
  corrections to W+ W- production via vector-boson fusion}},  {\em JHEP} {\bf
  07} (2006) 015, [\href{http://xxx.lanl.gov/abs/hep-ph/0603177}{{\tt
  hep-ph/0603177}}].

\bibitem{Alwall:2007st}
J.~Alwall, P.~Demin, S.~de~Visscher, R.~Frederix, M.~Herquet, {\em et.~al.},
  {\it {MadGraph/MadEvent v4: The New Web Generation}},  {\em JHEP} {\bf 0709}
  (2007) 028, [\href{http://xxx.lanl.gov/abs/0706.2334}{{\tt
  arXiv:0706.2334}}].

\bibitem{Martin:2009iq}
A.~D. Martin, W.~J. Stirling, R.~S. Thorne, and G.~Watt, {\it {Parton
  distributions for the LHC}},  {\em Eur. Phys. J.} {\bf C63} (2009) 189--285,
  [\href{http://xxx.lanl.gov/abs/0901.0002}{{\tt arXiv:0901.0002}}].

\bibitem{:2009je}
{\bf CDF and D0} Collaboration, {\it {Combined CDF and D0 Upper Limits on
  Standard Model Higgs-Boson Production with 2.1 - 5.4 fb**-1 of Data}},
  \href{http://xxx.lanl.gov/abs/0911.3930}{{\tt arXiv:0911.3930}}.

\bibitem{Campbell:2010cz}
J.~M. Campbell, R.~K. Ellis, and C.~Williams, {\it {Hadronic production of a
  Higgs boson and two jets at next- to-leading order}},  {\em Phys. Rev.} {\bf
  D81} (2010) 074023, [\href{http://xxx.lanl.gov/abs/1001.4495}{{\tt
  arXiv:1001.4495}}].

\bibitem{Klamke:2007cu}
G.~Klamke and D.~Zeppenfeld, {\it {Higgs plus two jet production via gluon
  fusion as a signal at the CERN LHC}},  {\em JHEP} {\bf 04} (2007) 052,
  [\href{http://xxx.lanl.gov/abs/hep-ph/0703202}{{\tt hep-ph/0703202}}].

\bibitem{Aad:2010ey}
{\bf Atlas Collaboration} Collaboration, G.~Aad {\em et.~al.}, {\it
  {Measurement of the top quark-pair production cross section with ATLAS in pp
  collisions at $\sqrt{s}=7$ TeV}},
  \href{http://xxx.lanl.gov/abs/1012.1792}{{\tt arXiv:1012.1792}}.

\bibitem{Khachatryan:2010ez}
{\bf CMS Collaboration} Collaboration, V.~Khachatryan {\em et.~al.}, {\it
  {First Measurement of the Cross Section for Top-Quark Pair Production in
  Proton-Proton Collisions at sqrt(s)=7 TeV}},  {\em Phys.Lett.} {\bf B695}
  (2011) 424--443, [\href{http://xxx.lanl.gov/abs/1010.5994}{{\tt
  arXiv:1010.5994}}].

\bibitem{Cacciari:2008gp}
M.~Cacciari, G.~P. Salam, and G.~Soyez, {\it {The anti-$k_t$ jet clustering
  algorithm}},  {\em JHEP} {\bf 04} (2008) 063,
  [\href{http://xxx.lanl.gov/abs/0802.1189}{{\tt arXiv:0802.1189}}].

\bibitem{Cacciari:2005hq}
M.~Cacciari and G.~P. Salam, {\it {Dispelling the N**3 myth for the k(t)
  jet-finder}},  {\em Phys.Lett.} {\bf B641} (2006) 57--61,
  [\href{http://xxx.lanl.gov/abs/hep-ph/0512210}{{\tt hep-ph/0512210}}].

\bibitem{Dittmar:1996ss}
M.~Dittmar and H.~K. Dreiner, {\it {How to find a Higgs boson with a mass
  between 155-GeV - 180-GeV at the LHC}},  {\em Phys. Rev.} {\bf D55} (1997)
  167--172, [\href{http://xxx.lanl.gov/abs/hep-ph/9608317}{{\tt
  hep-ph/9608317}}].

\bibitem{Giele:2009ui}
W.~Giele, Z.~Kunszt, and J.~Winter, {\it {Efficient Color-Dressed Calculation
  of Virtual Corrections}},  {\em Nucl.Phys.} {\bf B840} (2010) 214--270,
  [\href{http://xxx.lanl.gov/abs/0911.1962}{{\tt arXiv:0911.1962}}].

\bibitem{Bern:1991aq}
Z.~Bern and D.~A. Kosower, {\it {The Computation of loop amplitudes in gauge
  theories}},  {\em Nucl.Phys.} {\bf B379} (1992) 451--561.

\bibitem{Bern:2002zk}
Z.~Bern, A.~De~Freitas, L.~J. Dixon, and H.~Wong, {\it {Supersymmetric
  regularization, two loop QCD amplitudes and coupling shifts}},  {\em
  Phys.Rev.} {\bf D66} (2002) 085002,
  [\href{http://xxx.lanl.gov/abs/hep-ph/0202271}{{\tt hep-ph/0202271}}].

\end{thebibliography}
\end{document}